\documentclass{aa}
\usepackage{graphicx}
\usepackage{natbib}
\usepackage{latexsym}
\usepackage[american]{babel}
\bibpunct{(}{)}{;}{a}{}{,}

\def\app{Astropart. Phys.}
\def\lesssim{\mathrel{\hbox{\rlap{\hbox{\lower4pt\hbox{$\sim$}}}\hbox{$<$}}}}
\def\gtrsim{\mathrel{\hbox{\rlap{\hbox{\lower4pt\hbox{$\sim$}}}\hbox{$>$}}}}
\def\varP{\mathcal P}

\begin{document}

\title{Cosmic Rays from Microquasars:\\ a Narrow Component to the CR
Spectrum?}

\author{Sebastian Heinz\inst{1,3} \and Rashid Sunyaev\inst{1,2}}

\institute{Max-Planck-Institut f\"{u}r Astrophysik,
Karl-Schwarzschild-Str.~1, 85741 Garching, Germany \and Space Research
Institute (IKI), Profsouznaya 84/32, 117810 Moscow, Russia \and
\email{heinzs@mpa-garching.mpg.de}} \authorrunning{Heinz \& Sunyaev}
\titlerunning{A Narrow Component to the CR Spectrum}

\date{today}

\abstract{We propose that relativistic Galactic jets like those observed in
GRS 1915+105 and GRO J1655-40 may produce a small but measurable
contribution to the cosmic ray (CR) spectrum. If these jets contain cold
protons and heavy ions (as in the case of SS433), it is plausible that this
component will consist of a narrow spectral feature, with a mean particle
energy corresponding roughly to the bulk kinetic particle energy in the
beam, $\Gamma_{\rm jet}\,{\rm m_p}\, c^2$. Based on the current estimates
of $\Gamma_{\rm jet}$, this feature will fall into the range of 3--10
GeV. The presence of several sources with different $\Gamma_{\rm jet}$ will
lead to the superposition of several such peaks. In addition to the narrow
peaks, diffusive particle acceleration should also produce a powerlaw,
whose low energy cutoff at or above $\Gamma_{\rm jet}^2\,m_{\rm p}\,c^2$
would be visible as an additional spectral feature. The large metallicities
measured in several binary companions of jet sources suggest that this CR
component could have an anomalous composition compared to the bulk Galactic
CR spectrum.  We provide estimates of the effects of adiabatic losses which
are the greatest challenge to models of narrow band CR production in
microquasar jets. While the total energy contained in the microquasar CR
component is highly uncertain, the local CR spectrum in the vicinity of any
microquasar should be severely affected. The upcoming {\em AMS 02}
experiment will be able probe the low energy CR spectrum for such
components and for composition anomalies. The spectrally peculiar gamma ray
emission produced by interaction of the ISM with CRs surrounding
microquasars might be detectable by {\em GLAST}. If the presence of a
microquasar CR proton component can be ruled out observationally, this
argument could be turned around in favor of electron-positron jets.  We
show that existing {\em OSSE/GRO} and future {\em INTEGRAL} data on the
Galactic 511 keV line flux put interesting constraints on the particle
content of microquasar jets.  The process of CR production in relativistic
flows inside the Galaxy is fundamentally different from the standard
picture of CR production in nonrelativistic shocks in supernova remnants,
because the particles injected by a relativistic flow are already
relativistic, without any need for diffusive
acceleration.}

\maketitle

\section{Introduction}
During the past decade, a new class of relativistic jet sources has been
discovered in the Galaxy: Galactic superluminal radio sources (or
microquasars), exemplified by the prototypical sources GRS 1915+105 and GRO
J1655-40.  Radio outbursts from these transient X-ray binaries, containing
accreting black holes or neutron stars, are associated with ejections of
plasma at relativistic bulk velocities.  The bulk Lorentz factors of these
flows inferred from observations are of the order of $\Gamma_{\rm jet} \sim
3 - 10$, though exact observational determination of $\Gamma_{\rm jet}$ is
difficult and in any case it is unlikely that a universal value of
$\Gamma_{\rm jet}$ holds for all jets.  For reasons of simplicity and for
lack of better knowledge we will adopt a fiducial value of $\Gamma_{\rm
jet} \sim 5$ for {\em numerical examples}, with the {\em explicit}
understanding that Galactic jets most likely operate at a variety of
Lorentz factors, which are currently not well determined.  These ejections
can contain energies in excess of $10^{44}\, \rm{ergs}$
\citep{mirabel:94,hjellming:95,fender:99}, and they occur at a Galactic
rate of $\gtrsim few\ {\rm yr^{-1}}$.

A large part of the kinetic energy transported by these jets is transferred
into random, isotropic particle energy at the interface between the jet and
the ambient medium, the working surface.  Because the jets are
relativistic, {\it the particles leaving the working surface must a priori
be relativistic without any need for diffusice acceleration.  This
mechanism of accelerating relativistic cosmic ray (CR) particles is
fundamentally different from CR production in the non-relativistic shocks
of supernova remnants (SNRs)}, which provides the bulk of the Galactic CRs.

While the momentum gain for particles crossing a non-relativistic shock is
small (of order $\delta\,p/p \sim v/c$), the large momentum gains
encountered in relativistic shocks (of order $\delta \,p/p \sim \Gamma$,
where $\Gamma$ is the shock Lorentz factor) should lead to the formation of
distinct spectral features in the spectrum (see Sect.~\ref{sec:shock}).
Thus, unlike the CRs produced in SNRs, which follow a smooth powerlaw
spectrum, the CRs produced in relativistic flows, like those encountered in
microquasars, should show clearly distinguishable, and possibly narrow,
spectral features.

\begin{figure}[tbp]
\begin{center}
\resizebox{\columnwidth}{!}{\includegraphics{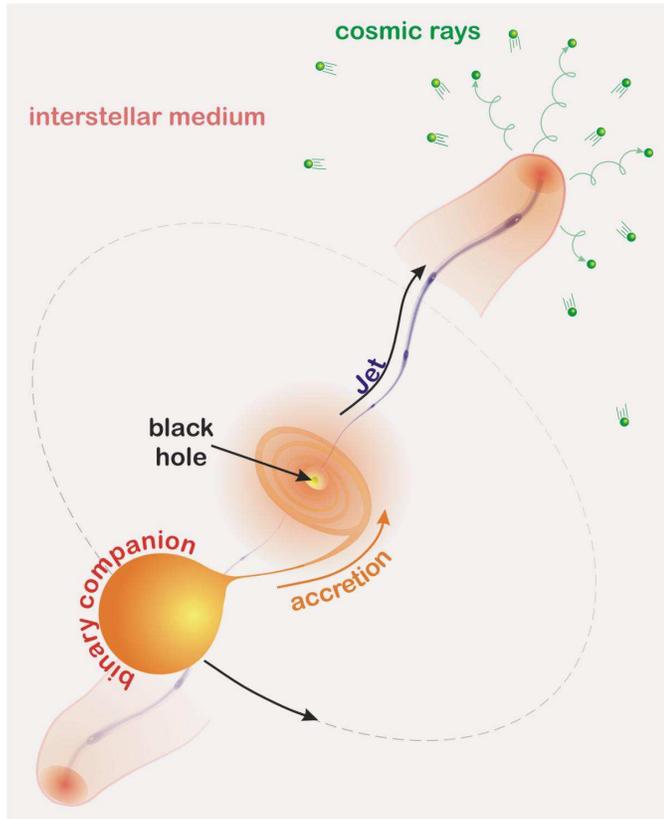}}
\end{center}
\caption{Cartoon of the proposed model of CR production in microquasars:
The interface between the relativistic jet and the ISM is a natural site
for the production and release of relativistic particles.
\label{fig:picture}}
\end{figure}

If these particles can escape the working surface without suffering
significant adiabatic energy losses, they will diffuse through interstellar
space, and will thus contribute to the Galactic cosmic ray (CR)
spectrum\footnote{The production of ultra-high energy CRs in Galactic X-ray
binaries has been discussed in the past in the context of Cyg X-3
\citep{lloyd-evans:83,manchandra:93,mitra:94}, though recent studies have
not confirmed the presence of ultra-high energy gamma rays
\citep{borione:97}.  This scenario is shown as a cartoon in
Fig.~\ref{fig:picture}.  The production of CRs in microquasars has so far
only received minor attention in the literature \citep[for example, it was
mentioned as an alternative source to produce all CRs by][]{dar:01}, and
requires further study.}.

Based on the arguments presented in this paper, we conclude that an
additional component of CRs generated by relativistic jets in microquasars
should exist in the Galaxy.  Initially, this component should consist of
narrow peaks, with peak energies corresponding to $\Gamma_{\rm jet}\,
m_{\rm p}\, c^2$ from different jet sources.

There are many mechanisms which might broaden these features.  However, any
observational limits on their existence would give us additional
information about the physics of microquasar jets and the physical
conditions in relativistic shocks.  Below, we will discuss the main
mechanisms which could smooth out the component under discussion.  In this
paper, we content ourselves with presenting order of magnitude estimates
only, since the goal of the paper is to point out to the CR community that,
in addition to CR acceleration in supernova remnants (SNRs), there is
another very effective mechanism to release relativistic particles in the
Galaxy.  Traces of these particles might be hidden in the observed CR
spectra, in $\gamma$-rays with energies of a few 100 MeV, and possibly in
electron-positron annihilation line emission from regions close to the
location of microquasars in the Galactic plane.

\section{Outline of the model}
In this section we will present a general description of the model proposed
for CR production in microquasars.

As we will argue below, it is likely that the plasma traveling far
downstream in the jet towards the working surface is cold (the mean
particle velocity is $\langle v^2\rangle \ll c^2$ in the rest frame of the
jet plasma), especially if microquasar jets are composed of electron-ion
plasma. The same is, of course, also true for the undisturbed interstellar
medium (ISM).  Thus, the bulk of the plasma transported to the interface
between the jet and the ISM (for simplicity we will call this interface the
working surface of the jet, regardless of its detailed physical structure)
is initially cold.

This conjecture is inspired by observations of the mildly relativistic jets
in SS433 (the best studied relativistic Galactic jet to date, albeit mildly
relativistic and not considered a microquasar).  In this source, red- and
blue-shifted Balmer H$\beta$ and other optical recombination lines, usually
radiated by plasmas with temperatures of order $10^{4}\,{\rm K}\,\sim
1\,{\rm eV}$, allow the determination of the bulk velocity of the flow:
$0.26\, c$.  This velocity is remarkably constant over the 20 years the
source has been observed \citep{margon:84,milgrom:82}.  {\em ASCA}
\citep{kotani:98} and recent {\em Chandra} \citep{marshall:02} observations
of X-ray lines of hydrogen- and helium-like ions of Iron, Argon, Sulfur,
and Oxygen show that these ions are moving in the flow with the same
velocity, $0.26\, c$.  This X-ray emitting plasma at temperatures of $T
\sim 10^{7}-10^{8}\,{\rm K} \sim 10^{3}-10^{4}\,{\rm eV}$ is observed at
much smaller distances ($\sim 10^{11}\,{\rm cm}$) from the central compact
object than the optical line emission region ($\sim 10^{14}\,{\rm cm}$).  A
striking feature of the SS433 jet is that the plasma is observed to be
moving with relativistic velocities. Yet, at the same time the jet plasma
itself shows very little line broadening (i.e., it is cold).

If microquasar jets are similar to the SS433 jets in composition and
properties (i.e., cold electron-ion plasma at relativistic bulk speeds) the
consequences for the interpretation of these jets will be far reaching, as
we will argue below.  Independent from this argument, the radio synchrotron
emission detected from microquasar jets and several radio nebulae
surrounding microquasar sources (see Sect.~\ref{sec:energetics}) is clear
evidence for the presence of relativistic electrons, which, when released
into the ISM, will act as cosmic ray electrons.

\subsection{The standard picture for jet working surfaces}
\label{sec:standard}

\begin{figure}[tbp]
\begin{center}
\resizebox{\columnwidth}{!}{\includegraphics{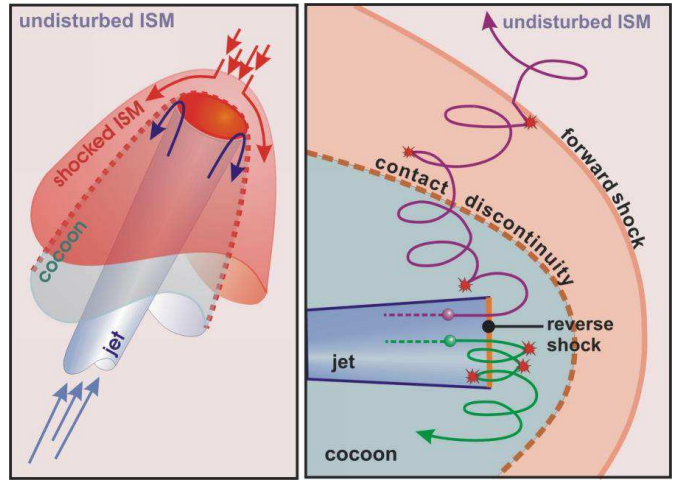}}
\end{center}
\caption{Cartoon of the standard picture of the interface between jet and
ISM ({\em left panel}), as envisaged to apply in FR II radio galaxies.  The
injection of relativistic particles can occur either in the reverse or
forward shock.  {\em Right panel}: cartoon of particle trajectories for
particles crossing the shock only once (upper solid line) and particles
participating in diffusive shock acceleration (lower solid line), particle
scattering indicated as stars.
\label{fig:cartoon}}
\end{figure}

\begin{figure}[tbp]
\begin{center}
\resizebox{\columnwidth}{!}{\includegraphics*{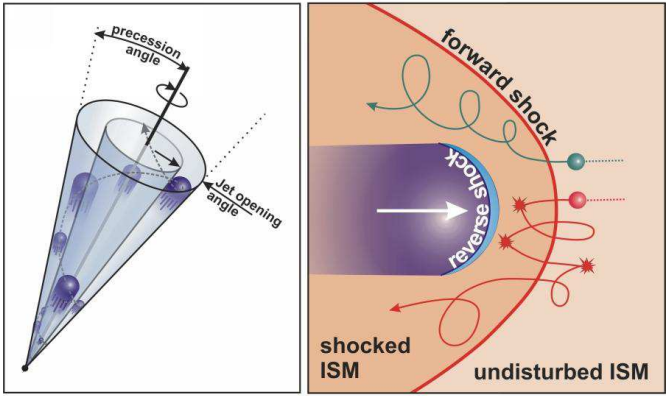}}
\caption{{\em Left:} Cartoon of a jet composed of discrete ejections with
precession.  In such a non-stationary picture, each ejection is slowed by
its interaction with the ISM (which might be disturbed by previous
ejections).  This interaction will likely happen in the form of a forward
shock (into the ISM).  ISM particles will leave the shock with energies of
order $\Gamma_{\rm jet}\,m_{\rm p}\,c^2$ (here, $\Gamma_{\rm jet}$ is the
Lorentz factor of the individual ejection, which decreases with time). {\em
Right:} Cartoon of particles crossing a relativistic forward shock, likely
the appropriate scenario for a cold ejection running into the ISM.  Same
nomenclature as in Fig.~\ref{fig:cartoon}.\label{fig:bullet_cartoon}}
\end{center}
\end{figure}

The standard picture for the interface between powerful radio galaxies and
their environment is a strong double shock structure (forward shock into
the ISM and reverse shock into the jet), shown in Fig.~\ref{fig:cartoon}.
The shocked jet material is shed at the head of the jet and inflates an
enshrouding cocoon around the jet, filled with relativistic plasma, which
has gone through the terminal shock.  Such a scenario might also be
relevant for the terminus of Galactic relativistic jets.  A similar picture
arises if the jets are composed of discrete ejections, propagating into an
external medium at relativistic speeds, as sketched in
Fig.~\ref{fig:bullet_cartoon}.

A cold upstream particle crossing an ultra-relativistic shock into a
downstream region with relative Lorentz factor $\Gamma_{\rm rel} \sim
\Gamma_{\rm jet}$ will have an internal energy of $\gamma m c^2 \sim
\Gamma_{\rm jet} m c^2$ in the downstream frame after the first shock
crossing.  Consequently, all initially cold particles will leave the shock
with about the same specific energy $\Gamma_{\rm jet} c^2$.  Particles can
pick up additional energy if they cross the shock multiple times, which is
the basis of diffusive shock acceleration schemes like Fermi acceleration,
resulting in the formation of a powerlaw distribution.  However, as has
recently been shown by \citet{achterberg:01}, the bulk of the particles
crossing a relativistic shock escape after the very first shock passage and
will therefore not participate in diffusive shock acceleration.  It is
these particles that carry off the bulk of the dissipated jet energy.

As a result, the bulk of the particles might leave the shock with a narrow
energy distribution, peaking at an energy close to the specific kinetic
energy of the jet: $\langle \gamma m c^2\rangle \sim \Gamma_{\rm jet} m
c^2$, with an energy width similar to or higher than the Lorentz
transformed thermal velocity, $\Delta \gamma \sim 2\,\gamma\,c_{\rm s}$
(i.e., very narrow, since the internal sound speed $c_{\rm s}$ is small:
$c_{\rm s} \ll 1$).

Whether this narrow distribution will be preserved as the particles travel
away from the shock, or whether it will be thermalized, depends on the
efficiency of collective plasma effects and small angle scattering on
magnetic field irregularities, which are also needed to isotropize the
particle distribution.  If collective effects are strong, the particle
spectrum will be broadened into a relativistic Maxwell-Boltzmann
distribution, with a temperature corresponding to the value given by the
relativistic Rankine-Hugoniot jump conditions.  In the case of a strong,
ultra-relativistic shock, this is simply $kT \sim 1/3 \Gamma_{\rm jet} m
c^2$, i.e., the mean particle energy is just $\Gamma_{\rm jet} m c^2$
\citep[e.g.][]{blandford:76}.  In this case the relativistic proton plasma
in the shocked ISM is equivalent to the X-ray emitting gas in SNR shocks.
However, in microquasars shocks we have extremely rarefied, relativistic
particles with a relatively narrow thermal (i.e., not powerlaw) energy
distribution.

\subsection{Termination without strong shocks?}
\label{sec:diffuse}
\begin{figure}[tbp]
\begin{center}
\resizebox{0.85\columnwidth}{!}{\includegraphics{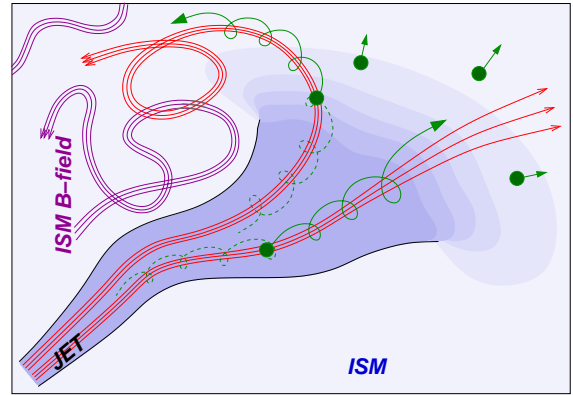}}
\end{center}
\caption{Cartoon of particle injection by a jet without a strong shock at
its interface with the ISM (e.g., if the magnetic structure of the jet is
connected with the ISM and the jet is sub-Alfv$\acute{\rm e}$nic).  Dots
represent CRs transported by the jet and released into the ISM at the bulk
speed of the jet, grey arrows represent magnetic flux tubes.  The right
flux tube representation depicts the case where adiabatic losses are
important, the left shows a case where the field flux tube is smoothly
fused with the stochastic interstellar field and adiabatic losses might not
be dominant in particle transport.
\label{fig:diffuse_cartoon}}
\end{figure}

However, the structure of relativistic shocks is still not well understood
and it might be that this interface is not a simple double shock structure.
It could be significantly different in nature. For example, the jet could
be magnetically connected with the environment, i.e., if the flux tubes
join smoothly with the large scale magnetic field of the ISM, as shown in
the cartoon in Fig.~\ref{fig:diffuse_cartoon} (note, however, that
\citealt{lubow:94} showed that realizing such configuration is rather
difficult).

In such a case the shock would be replaced by stochastic pitch angle
scattering of the particle distribution (this can occur if the jet is
moving sub-Alfv$\acute{\rm e}$nically, for example).  Since the jet plasma
is traveling relative to the ISM, such a scenario would excite strong
two-stream instabilities at the interface between ISM and jet plasma, which
would isotropize and possibly thermalize the particle distribution of the
jet very quickly.  The deposition of jet thrust would then imply that this
interface is itself moving through space.  Precession, as observed in SS433
\citep[e.g.,][]{milgrom:79} and suggested to be present in GRO 1655-40
\citep{hjellming:95}, will significantly alter the dynamical balance
between ISM and jet plasma, as will the time dependent nature of the
interface if the jets are composed of discrete ejections.

If furthermore the magnetic field is stochastically tangled on small
scales, the detailed behavior of the plasma could be very complicated, with
a gradual change from relativistic, ballistic motion to random propagation.
Qualitatively, this would be comparable to extragalactic FR I sources
(though the exact nature of the dynamics in FR I sources is not yet clear,
either).

In such a case, the absence of a strong shock would preclude diffusive
shock acceleration (though stochastic acceleration might still exist if
particles scatter off of relativistic turbulence which might exist in the
transition region between jet and ISM). Only the narrow or thermalized
component with mean energy of $\Gamma_{\rm jet} m_{\rm p} c^2$ and strong
cutoff at higher energies would exist.

\subsection{Spectral characteristics}
\label{sec:spectrum_intro}
Based on Sect.~\ref{sec:standard} and Sect.~\ref{sec:diffuse}, we therefore
predict that each Galactic microquasar produces a narrow component of CRs,
which peaks at an energy $\Gamma_{\rm jet} m c^2$.  For protons, this
energy should fall into the range of $\Gamma_{\rm jet} m_{\rm p} c^2 \sim 3
- 10$ GeV.  The superposition of these spectral signatures from several
microquasars will appear as a broad feature in the energy range from $3 -
10$ GeV.

\begin{figure*}[tbp]
\begin{center}
\resizebox{0.99\columnwidth}{!}{\includegraphics{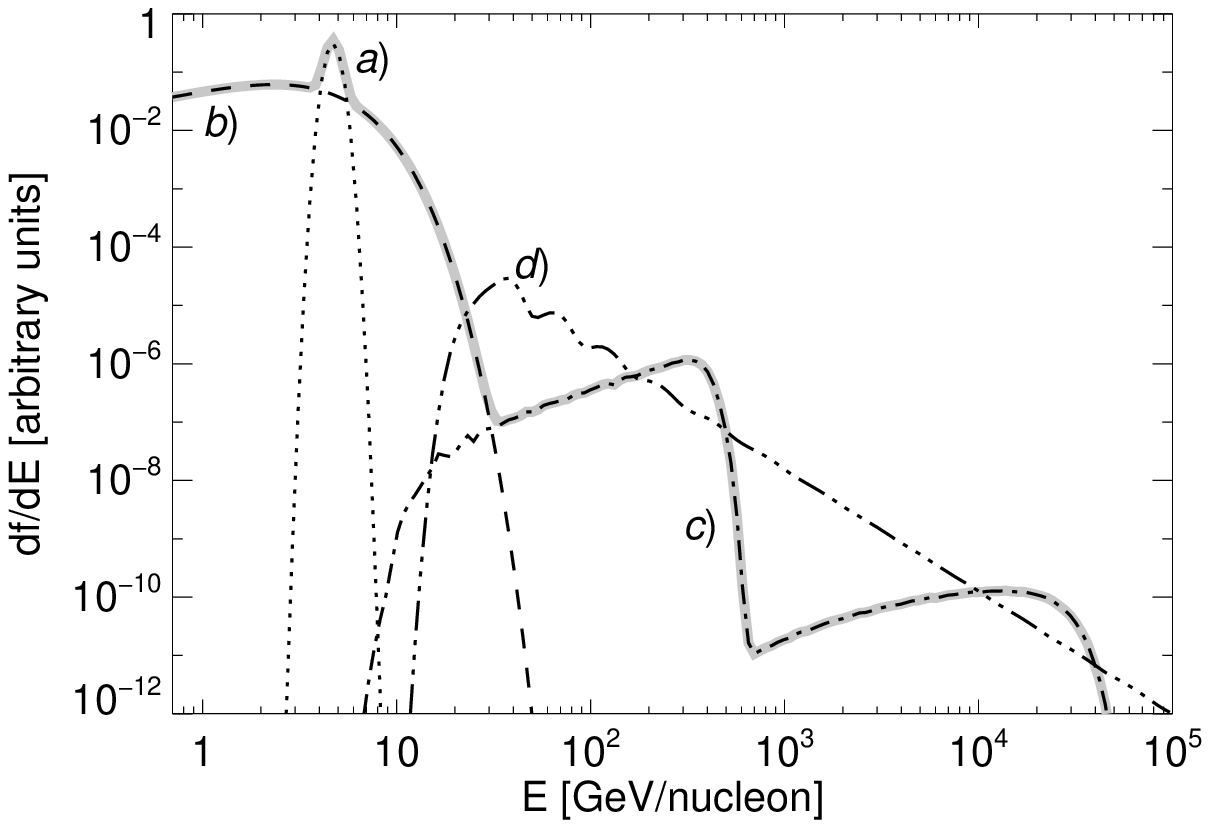}}
\resizebox{0.99\columnwidth}{!}{\includegraphics{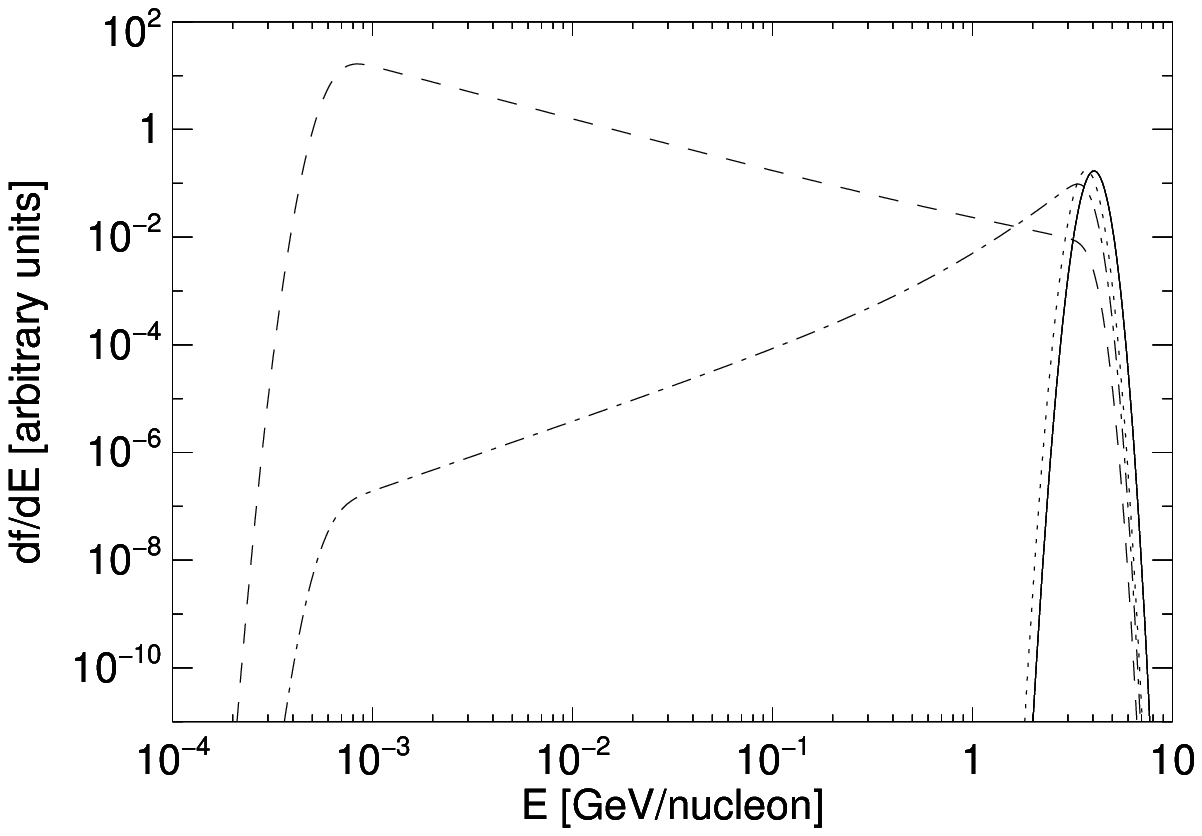}}
\end{center}
\caption{{\em Left Panel}: Sketch of the predicted contribution from a
microquasar to the Galactic CR spectrum for $\Gamma_{\rm jet}=5$.  {\em a)
Dotted line}: narrow component, corresponding to cold ($T\sim
7\times10^{10}\,{\rm K}$) upstream particles simply isotropized upon
crossing the working surface (Lorentz transformed energy distribution),
arbitrary normalization; {\em b) dashed line}: Maxwell-Boltzmann component
from particles thermalized in the shock (normalized to contain the same
power as component {\em a}); {c) \em dash-dotted line}: powerlaw-like
component (escape probability 99\%, normalized to contain 10\% of component
{\em a}) for the case of strong scattering.  Note the similarity to
Comptonization of low energy seed photons by a relativistic medium with low
optical depth (i.e., high escape probability); {\em d) dash-triple-dotted
line}: powerlaw spectrum for weak scattering case \citep{achterberg:01} for
the same normalization as component {\em c}).  Note the difference in the
position of the first peak between {\em c)} and {\em d)}, $\sim \Gamma_{\rm
jet}^2$ vs.  $\sim \Gamma_{\rm jet}^3$.  {\em Right Panel}: estimate of the
effects of adiabatic losses on the particle distribution for different
ratios of the adiabatic losses time $\tau_{\rm ad}$ of the particles to the
escape time $\tau_{\rm esc}$, compared to the injected particle
distribution (solid) line for a pressure differential between shock and ISM
of $p_{\rm shock}/p_{\rm ISM}=3\times 10^8$ (see Appendix
\ref{sec:appendix-diffusion}): strong adiabatic losses ($\tau_{\rm
ad}/\tau_{\rm esc} = 10^{-3}$, dashed line), intermediated losses
($\tau_{\rm ad}=\tau_{\rm esc}$, dash-dotted line), and weak adiabatic
losses ($\tau_{\rm ad}/\tau_{\rm esc}=10^{3}$, dotted line).
\label{fig:sketch}}
\end{figure*}

Even if most of the particles are thermalized downstream, the spectrum will
still show a steep turnover beyond energies of order $3kT \sim \Gamma_{\rm
jet}\, m_{\rm p}\, c^2$ (see the dashed curve in Fig.~\ref{fig:sketch}),
which will appear as an edge-like feature in the overall CR spectrum.

Similarly, a number of other processes will tend to broaden any narrow
component produced in the working surface, including adiabatic losses
(competing with diffusion of particles out of the loss region, see Appendix
\ref{sec:appendix-diffusion} and the insert in Fig.~\ref{fig:sketch}) and
solar modulation.  The effect of these processes will be to spread
particles to lower energies, leaving the strong turnover/cutoff above
energies of $\Gamma_{\rm jet}\,m\,c^2$ intact.

The only serious cooling these CR protons at energies of a few GeV might
experience will be adiabatic losses, which will occur if the particles are
confined to an expanding plasma volume (e.g., if it is overpressured with
respect to the environment).  However, since many processes can lead to
increased diffusion of these particles, it appears plausible that a large
fraction of the CRs might escape before they suffer strong adiabatic
losses.

If a component of cold electrons is also present in the jets in addition to
the {\em observed} powerlaw electrons, a similar, {\em very low energy
relativistic electron component} (around 2 - 5 MeV) might appear.  However,
it would contain only a fraction $m_{\rm e}/m_{\rm p}$ of the energy in the
proton component.  

The remaining fraction of particles (both ions and electrons) which do not
escape the shock after the first shock passage and thus perform multiple
shock crossings will be accelerated diffusively to a powerlaw-like
distribution.  Only the high energy tail of this powerlaw-like electron
component is directly observable via synchrotron radio emission.

Likening the acceleration of particles crossing a relativistic shock to the
problem of Compton up-scattering of low energy photons on {\em relativistic
thermal} electrons \citep[see, for example,][]{pozdnyakov:83}, we note that
a particle scattered both up-stream and down-stream of the shock will
experience an energy gain by a factor of order $\Gamma_{\rm rel}^2$ per
crossing cycle, as was argued by, where $\Gamma_{\rm rel} \sim \Gamma_{\rm
jet}$ is the relative Lorentz factor between upstream and downstream
plasma.  This was argued by \cite{vietri:95}, applied to the acceleration
of particles in gamma ray burst shocks.  This will lead to the production
of several peaks in the spectrum.  The input spectrum for this
up-scattering process is the narrow particle population produced in the
initial shock crossing (discussed above), and thus peaks will appear at
energies $\sim few\times{\Gamma_{\rm jet}}^{2i + 1}\, m_{\rm p}\, c^2$,
where $i$ is the number of shock crossing cycles performed by the particle.
The normalization of each peak, and thus the approximate powerlaw index, is
determined by the escape probability of the particles (similar to the
optical depth in inverse Compton scattering).  The resulting spectrum is
sketched in Fig.~\ref{fig:sketch}.

Note, however, that \citet{achterberg:01} argue that higher order shock
crossings do not lead to energy gains of order $\Gamma_{\rm rel}^2$.  In
their treatment, scattering is limited to very small angles and the energy
gain is only of order unity, and thus the position of the peaks would be
much more closely spaced, resembling a powerlaw much more than in the
Compton scattering analogy discussed in the previous paragraph.  The low
energy turnover (or cutoff) of this powerlaw distribution would then be
located roughly at $\Gamma_{\rm jet}^2 \sim 10 - 100$ GeV.  At higher
energies, multiple scattering will form a powerlaw with index\footnote{Note
that powerlaw indices from cosmic ray modified shocks tend to be steeper at
least near the low energy cutoff \citep[e.g.,][]{berezhko:99,ellison:02}}
$s \sim 2.3$.  According to this simple approach, the difference between
these two pictures is therefore the energy of the second peak ($\sim
\Gamma_{\rm jet}^3$ vs. $\sim \Gamma_{\rm jet}^2$).

Since the structure of relativistic shocks, and their presence in the
working surfaces of microquasar jets are subject to considerable
uncertainty, the observational discovery of any of the features discussed
in this paper (and in particular the second peak, which would help to
distinguish between the two scenarios of diffusive acceleration mentioned
in the previous paragraphs, see Fig.~\ref{fig:sketch}) or evidence of their
absence would be important input into theories of relativistic shocks.

\subsection{Comparison with the canonical CR powerlaw component}
The CR components described above and shown in Fig.~\ref{fig:sketch} should
be compared to the well known powerlaw CR component observed near earth: CR
protons and nuclei have a powerlaw spectrum with a uniform slope around $s
\sim 2.5 - 2.7$ over an extremely broad energy range from 1 GeV up to
$10^{15}$ eV.  Locally, the CR energy density is of order $10^{-12}\,{\rm
ergs\,cm^{-3}}$.  Observations of the light elements produced by spallation
reactions show that the lifetime $\tau_{\rm CR,\,disk}$ of relativistic
protons in the Galactic disk is of order $\tau_{\rm CR,\,disk} \sim 1.5
\times 10^{7}\,{\rm yrs}$ \citep{yanasak:01}, while the lifetime in the
Galactic halo $\tau_{\rm CR,\,halo}$ is close to $10^8$ years
\citep[e.g.,][]{ginzburg:96}.  This enables us to estimate the CR
luminosity of the Galaxy
as $L_{\rm CR} \sim 4 \times 10^{40}\,{\rm ergs\ s^{-1}}$, with a diffusion
coefficient $\kappa$ inside the disk of
\begin{equation}
	\kappa \sim \frac{H_{\rm disk^2}}{\tau_{\rm CR,\,disk}} = 2\times
	10^{28}\frac{\rm cm^2}{s}\frac{H_{\rm kpc}^2}{\tau_{15}} \sim
	10^{28.3}\,\frac{\rm cm^2}{\rm s}\,\kappa_{28.3}
\end{equation}
where $H_{\rm kpc}$ is the disk height in kpc, $\tau_{15}$ is the CR
lifetime in the disk in units of 15 Myrs, and $\kappa_{28.3}$ is the ISM
diffusion coefficient, normalized to a value of $2\times 10^{28}\,{\rm
cm^2\,s^{-1}}$.  The current paradigm for the bulk of the CRs observed in
the vicinity of the earth is that they are produced by shock acceleration
in the decelerating blast waves of SNRs \citep[e.g.][]{blandford:78}.  The
usual assumption is that about 5\% of the mechanical energy of the SNR are
converted into CR energy.

There is no doubt that in the vicinity of an active microquasar the low
energy part of the Galactic CR spectrum must be strongly distorted.  As a
result, smooth maxima or edge-like features should exist in the $few$ GeV
range of the CR spectrum.  For a distant observer, the signals from several
sources will be superimposed due to the long diffusion time through the
galaxy.  Integrally, though, deviations from the powerlaw spectrum expected
in diffusive shock acceleration models should be observable.

Energy estimates which we present below show that this CR component
produced in microquasars might contribute measurably to the spectrum of the
CR protons in the energy band mentioned above. We will argue that,
globally, microquasars should contribute upward of 0.1\% of the total
Galactic CR power. However, the locally measured (i.e., near earth)
relative strength of the proposed CR components produced in microquasars
compared to the canonical CR powerlaw distribution is highly uncertain, as
it depends on the history of microquasar activity in our Galactic
neighborhood.

Given these uncertainties, it might be rather difficult to detect the tiny
deviations in the CR spectrum caused by distant microquasars (further
complicated by the strong effects of solar modulation at and below the
predicted energy range). However, they might be measurable by the upcoming
{\em AMS 02} experiment \citep[e.g.][]{barrau:01}, which will offer
unprecedented sensitivity and will be launched during the upcoming solar
minimum (reducing the effects of solar modulation significantly).  Traces
of such a component might also be present in already existing high quality
data sets from past or ongoing experiments, such as {\em IMAX}
\citep{menn:00} or {\em CAPRICE} \citep{boezio:99}.

Absence of any traces of spectral deviations in the upcoming {\em AMS 02}
experiment might become a strong argument in favor of electron-positron
jets in Galactic superluminal radio sources or, alternatively, it would
demonstrate that there is an unknown acceleration mechanism with 100\%
efficiency of transforming of the mechanical beam energy into a
relativistic powerlaw distribution.  Given these premises, we can state
that one of the following two statements must hold: {\em 1)} Either an
additional hadronic CR component exists (though it may be so weak that
detection inside the solar system is impossible) or {\em 2)} All jets are
electron-positron dominated (in which case an additional CR
electron-positron component should exist).

\subsection{Peculiar abundances}
Recent studies of the chemical composition of Galactic transients indicate
that the binary companions of several microquasar sources are metal
enriched: the heavy element abundances (e.g., N, O, Ca, Mg) in the optical
counterparts of the X-ray sources GRO J1655-40 \citep{israelian:99} and
V4641 Sgr \citep{orosz:01} exceed solar abundances by about an order of
magnitude, much more so than the observed overabundance of the same
elements in CRs \citep{zombeck:90}.  Furthermore, the relative abundances
between these elements are unusual compared to either solar abundances or
the abundance pattern observed in the bulk of the CR spectrum.

These abundance anomalies in GRO J1655-40 and V4641 Sgr could be the result
of mass exchange between two rapidly evolving massive stars or enrichment
of the normal stellar atmosphere during the supernova explosion of the
primary predecessor.  Accretion brings these abundance anomalies into the
jet creation region in the inner disk, from where they could be transported
out by the jet, eventually producing CRs by the mechanism outlined above.
Similarly, Cyg X-3 is known to have an extremely hydrogen deficient
Wolf-Rayet companion \citep{vankerkwijk:92,vankerkwijk:96,fender:99b},
which could also lead to a large overabundance in helium and heavier
elements relative to hydrogen in the produced CR spectrum.

Therefore, Galactic jets might be responsible for part of the observed CR
abundance anomalies.  Moreover, the CR component produced in relativistic
jet sources inside the Galaxy might show rather unusual chemical
abundances, in comparison with the bulk of the CRs in the powerlaw
population.  This would immediately distinguish Galactic jets from other CR
creation mechanisms.  The comparison between the measured abundances in the
energy range where we expect Galactic jet sources to contribute (of order a
few GeV) with those measured in the pure powerlaw regime will thus be an
important probe to search for the proposed CR component.  Note that the CRs
produced in SNRs originate in the external shock of the swept up ISM, thus
the abundances of the produced CR spectrum reflect the ISM, which might
have been enriched by a pre-collapse wind, but will not show the peculiar
abundance of the SN ejecta.  Because all the spectral components
accelerated in microquasars originate from the same plasma, they should all
show the same abundance pattern.  This could be a way to associate spectral
features at different CR energies with a microquasar origin.

A second way to distinguish particles accelerated in the relativistic
shocks of Galactic microquasars from those accelerated in non-relativistic
SNR shocks is the different energy---particle mass relation: All particles
in relativistic cold jets have the same Lorentz factors.  Since single-pass
shock acceleration will accelerate all particles to roughly the same random
Lorentz factor, the peak energy for different species will be proportional
to their rest mass (i.e., a fixed energy per nucleon).  Electromagnetic
acceleration processes would instead produce particle energies proportional
to $Z/A$.  This difference might again be measurable by {\em AMS 02}, and
might already be present in CR data on heavy nuclei from experiments like
{\em HEAO-3} \citep{engelmann:90} or {\em ACE} \citep{binns:01}.

\section{Energetics and spectra of microquasar CRs}
\label{sec:acceleration}

\subsection{Microquasar energetics}
\label{sec:energetics}
The best studied cases of microquasar activity show that large amounts of
kinetic energy can be liberated: Conservative equipartition estimates of
the energy released in the major outbursts of GRS 1915+105 give $E_{\rm
kin} > 2 \times 10^{44}\,{\rm ergs}$ \citep{mirabel:99,fender:99}, released
over a period of a few days at most.

Existing radio monitoring data \citep{pooley:97,foster:96} show that GRS
1915+105 exhibits several giant flares per year, not all of which were
observed with detailed campaigns \citep[e.g.,][]{fender:99}.  This yields
an estimated average kinetic power, and, since almost all of the energy
will initially be deposited in the form of CRs, an estimated cosmic ray
power of $L_{\rm kin} > 10^{37}\,{\rm ergs\ s^{-1}}$ for GRS 1915+105
alone.  In fact, GRS 1915+105 seems to release an even higher power in the
form of microflares between major outbursts, estimated to exceed ${\rm few}
\times 10^{37}\,{\rm ergs\ s^{-1}}$ \citep{mirabel:99} and even $3\times
10^{38}\ {\rm ergs\ s^{-1}}$ \citep{fender:00b}.

\begin{figure*}
\begin{center}
\resizebox{0.9\textwidth}{!}{\includegraphics{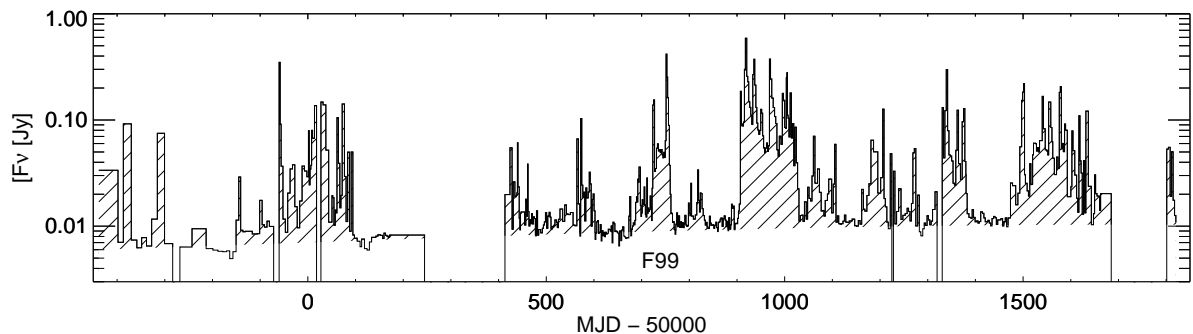}}
\end{center}
\caption{Plot of the 2.25 GHz GBI monitoring data ({\tt
http://www.gb.nrao.edu/fgdocs/gbi/gbint.html}) for GRS 1915+105 over the
time span from June 1994 to August 2000.  Shown as a hatched area is the
flux considered above the baseline flux, i.e., the flux considered to
originate from flares, which we integrated to arrive at the estimate for
the average kinetic power carried by the jet.  The flare analyzed by
\cite{fender:99} is indicated by the mark ``F99''.  The large gaps are due
to gaps in the monitoring campaign and were not included in the procedure.
\label{fig:gbi}}
\end{figure*}

Using the publicly available GBI monitoring data ({\tt
http://www.gb.nrao.edu/fgdocs/gbi/gbint.html}), we estimate that GRS
1915+105 spends in excess of 60\% of its time at flux levels significantly
enhanced over the baseline flux (GBI monitoring data of Cyg X-3 show a
similar rate), with about 2 major outbursts per year \citep[see
Fig.~\ref{fig:gbi} and also][]{fender:99,foster:96}.  Assuming that the
observed radio flux in flares is proportional to the amount of kinetic
energy released in the flare, and using the observed 1997 flare
\citep{fender:99} with a minimum kinetic energy estimate of $2\times
10^{44}\,{\rm ergs}$, we estimate that over the period covered by GBI
monitoring, the mean kinetic luminosity of GRS 1915+105 in flares is of
order $L_{\rm kin} \sim 10^{38}\,{\rm ergs\,s^{-1}}$.  If the baseline
radio emission from GRS 1915+105 is also due to low level jet emission,
this estimate increases by a factor $\sim 1.4$.

The jets in SS433 are even more impressive: Reasonable estimates put the
total, continuous kinetic power in excess of $L_{\rm kin} \sim few \times
10^{38} - few \times 10^{39}\ {\rm ergs\ s^{-1}}$
\citep{marshall:02,margon:84,spencer:84}, which is already of order 1 -
10\% of the total Galactic CR luminosity.  While the jets in SS433 are only
mildly relativistic, and the production of observable, relativistic CRs
thus falls under similar restrictions with regard to particle acceleration
efficiency as supernovae, the example of SS433 does show that Galactic jets
are capable of releasing impressive amounts of kinetic energy.  Thus, jets
from objects like SS433 (with mildly relativistic bulk speeds) might be an
important source of sub-CRs in the Galaxy, influencing heating and
ionization of the ISM.

Since the subject of Galactic microquasars is still relatively young, and
many of the known sources have only been discovered in recent years,
estimating the true Galactic rate of radio outburst events and thus the
total Galactic power in relativistic jets is difficult.  Taking the
interval from 1994 through 2000, there were at least 7 well observed giant
radio outbursts comparable in strength with GRS 1915+105 (corrected for
Galactic distance) in the sources Cyg X-3 \citep{mioduszewski:01}, GRO
J1655-40 \citep{hjellming:95}, GRS 1915+105 \citep{mirabel:94,fender:99},
V4641 Sgr \citep{orosz:01}, and XTE J1748-288 \citep{fender:01}, giving a
very conservative lower limit on the Galactic rate of $1 \ {\rm yr}^{-1}$.
As with GRS 1915+105, we expect many giant flares to have gone unnoticed,
and a more reasonable estimate of the event rate would be of the order of
10-100 Galactic events per year.

Cyg X-3, which is believed to be relativistically beamed, shows radio peak
luminosities up to 200 times stronger than GRS 1915+105 \citep{fender:01},
and often exhibits flaring activity\citep{ogley:01} at or above the peak
level of GRS 1915+105 on timescales of $\sim 10$ days.  The other sources
mentioned above are very similar to each other in peak radio luminosity
\citep{fender:01}, which we take as an indicator of kinetic power (most of
these sources are not resolved and an estimate of the equipartition energy
of the jet is thus not possible).

If indeed these sources operate on the same level as GRS 1915+105, the
minimum kinetic luminosity of these seven sources together would be $L_{\rm
kin} \gtrsim 10^{39}\,{\rm ergs\,s^{-1}}$.  Since this estimate is based on
the minimum energy estimates of $L_{\rm kin}$ in GRS 1915+105, the true
kinetic luminosity of these sources might well be much larger.

Furthermore, there are many sources that are known to have been active at
earlier epochs [e.g., V404 Cyg \citep{han:92} and Cir X-1,
\citep{haynes:78}], which exhibited flux levels comparable to the above
mentioned sources.  Many sources currently active might simply not have
been detected yet.  Similarly, many more X-ray sources are observed to be
consistently active at lower radio fluxes (e.g., Cyg X-1, or GX 339-4,
\citealt{fender:01c}) than the brightest sources mentioned above.

During the past few years, it has become clear that radio emission from
Galactic X-ray sources is a very common phenomenon.  Radio emission is
usually detected during state changes of the X-ray source (into or out of
the low/hard state), including soft X-ray transients.  While the powerful
radio flares discussed above are associated with such transients, there are
many more X-ray sources which are active at lower radio flux levels (e.g.,
Cyg X-1, or GX 339-4, \citealt{fender:01c}).

These sources are observed to produce stationary, optically thick jet
emission (as opposed to the already optically thin emission detected in
typical radio flares of transient jets). It is not clear whether these jets
are in fact relativistic and how much energy they carry.  One might hope to
estimate the kinetic power from the observed flux, scaling it to the peak
flux observed in GRS 1915 as was done above for transient sources, but
detailed kinematic modeling of the jet would be required to justify such a
simple argument.  In any event, because no complete sample of such sources
exists, it is impossible to estimate the total fraction of mechanical jet
power contained in low power sources.  Furthermore, these sources might
have shown transient activity in the past as well, given that GRS 1915+105
also shows steady, quiescent radio emission at comparable flux levels.

Other sources like 1E~1740-294 \citep{mirabel:92}, GRS 1758-258
\citep{marti:98}, and Cir-X1 \citep{stewart:93} show permanent extended
structure resembling radio lobes in extragalactic radio sources, which are
witness to past radio activity and must harbor a significant amount of CRs
as well.  We note here that estimating the total kinetic power from the
presence or absence of radio lobes in microquasar sources (indicating past
activity) is severely hampered by the fact that Galactic radio lobes are
expected to have very low surface brightness \citep{heinz:02b}.

The estimate for the kinetic energy output from GRS 1915+105
\citep{fender:99}, which we used as a template case to estimate the total
Galactic energy in jets, is based on the assumption that the jet plasma is
composed of a powerlaw of relativistic electrons and cold protons (for
charge conservation).  The electron spectrum was assumed to extend only
over the range in frequency observed in the radio.  While an IR detection
of the jet indicates a high energy tail of the spectrum
\citep{sams:96,mirabel:96,eikenberry:98}, a low energy component (down to
$\gamma \sim 1 - 10$) has never been observed in any jet due to lack of
viable emission mechanisms to reveal such a component.  The possibility of
detecting this component via inverse Compton scattering has been discussed,
for example, by \citet{ensslin:01}.

Finally, the physical structure of microquasar jets is still not known -
they could be made of either discrete ejections or a continuous stream of
matter.  If the jet is not composed of discrete ejections, but instead is a
continuous outflow with knots corresponding to internal shocks,
\citet{kaiser:00} have shown that, in the case of GRS 1915+105, the
estimate of the total kinetic energy carried in the jet (and thus the total
CR energy released in the working surface) is a factor of 10 higher than
the above estimate (though the instantaneous kinetic power is reduced),
corrections for the low energy end of the particle distribution
notwithstanding.

All of this indicates that the  lower limit of $L_{\rm kin} \sim few \times
10^{38}\  {\rm ergs\ s^{-1}}$  is conservative,  and it  might be  that the
kinetic luminosity from  microquasars is of the order of  10\% of the total
Galactic  cosmic  ray power  $L_{\rm  CR}$.   Clearly,  the uncertainty  in
frequency  and power  of radio  flares in  microquasars  warrants continued
monitoring of these sources to  answer the question of how important energy
input by these sources really is.

\subsection{The termination shock}
\label{ref:shock}
In the following, we will assume that a standard working surface as
depicted in Fig.~\ref{fig:cartoon} exists.  The conversion of kinetic to
internal energy will then take place either in the forward or the reverse
shock.

\subsubsection{Discrete ejections}
If the jets are composed of discrete ejections, then each ejection will
eventually convert its kinetic energy into some form of internal energy via
interaction with the ISM, not unlike the external shock encountered in GRB
afterglows.  During the phase of relativistic propagation of the ejection,
either the forward or the reverse shock (or both) must be relativistic.  It
is inside of the relativistic portion of the shock that most of the energy
is dissipated.

For a cold ballistically expanding ejection, most of the energy is
dissipated in the forward shock.  This is implied by the small opening
angle $\theta$ of the ejection: as long as $\theta \sim c_{\rm s}/
\Gamma_{\rm jet} c \ll 1/\Gamma_{\rm jet}$ (with $\Gamma_{\rm jet}$ being
the bulk Lorentz factor of the ejection), the characteristic transverse
size of the ejection $R$ is always much smaller than the distance $d$ over
which it slows down, as seen in the frame of the blob: $R \sim \theta d \ll
d/\Gamma_{\rm jet}$ (here, the factor of $1/\Gamma_{\rm jet}$ accounts for
the Lorentz contraction in going to the frame of the ejection).  Thus, the
deceleration time $d/\Gamma_{\rm jet} c$ is much longer than the light
crossing time of the ejection $R/c$, and the deceleration must occur in a
sub-relativistic shock.  This implies that the ejection is not heated to
relativistic temperatures, while the forward shock must be relativistic,
with shock velocity corresponding to $\Gamma_{\rm jet}$.

The particles passing through this shock will have energies of order
$\Gamma_{\rm jet}\,m_{\rm p}\,c^2\ \leq\ \Gamma_{\rm jet,0}\,m_{\rm
p}\,c^2$, where $\Gamma_{\rm jet,0}$ is the initial Lorentz factor of the
ejection (before interaction with the ISM).  Because the ejection is slowed
down, particles of a spread in energies are created in the forward shock,
though the spectrum will show a sharp turnover or cutoff at energies $e
\gtrsim \Gamma_{\rm jet,0}\,m_{\rm p}\,c^2$ (below this cutoff, it is
plausible that the spectrum rises with $dN/d\gamma \propto \gamma^{-3}$,
see Appendix\,\ref{sec:appendix-slow-down}).

\subsubsection{Continuous jets}
However, there are reasons to believe that at least in the well studied
cases of GRS 1915+105 \citep[][, though this question is up for
debate]{kaiser:00} and in SS 433, as well as in the persistent radio
outflows of low-hard state sources like GX 339-4 and Cyg X-1 the jets are
continuous streams of matter.  In which part of the shock the dissipation
occurs in a continuous jet depends on the jet thrust and the external
density: The advance speed of the jet is roughly given by ram pressure
balance.  The transition from forward to reverse shock dissipation then
occurs when
\begin{equation}
	\frac{L}{A_{\rm shock}\, \rho_{\rm x}\, c^3} = \frac{2 \,
	L_{38.5}}{l_{16}^2\, \theta_{0.1}^2\, n_{\rm ISM}} \sim 1,
	\label{eq:shock}
\end{equation}
where $L_{38.5}\equiv L_{\rm kin}/10^{38.5}\ {\rm ergs\ s^{-1}} \sim L_{\rm
kin}/3\times 10^{38}\ {\rm ergs\ s^{-1}}$ is the kinetic power,
$l_{16}=l_{\rm jet}/10^{16}\ {\rm cm}$ is the jet length, $A_{\rm shock}$
the surface area of the shock, $\theta_{0.1}\equiv \theta/0.1$ the jet
opening angle in units of 0.1 radian or 5.7$\circ$, and $n_{\rm ISM} \equiv
\rho_{\rm x}/m_{\rm p}$ the ISM particle density in units of $1\ {\rm
cm^{-3}}$.  Note that the working surface propagates, so $l_{16}$ is a
measured quantity, given by the age of the source, its power, and the ISM
density.  Strictly speaking, the jet thrust will have to be measured in the
shock frame, and the relativistic shock jump conditions applied, but in
order to estimate which of the two shocks is going to be relativistic, the
above approximation is sufficient.

The reverse shock is relativistic if the ratio on the left hand side of
Eq.~(\ref{eq:shock}) is smaller than unity, the forward shock is
relativistic if the ratio on the left hand side is larger than unity.  For
Galactic sources both cases can occur for appropriate external densities,
depending on the length of the jet.

\subsection{Particle acceleration in relativistic shocks}
\subsubsection{Nucleon acceleration}
\label{sec:shock}
For non-relativistic shocks, it has long been known that multiple shock
crossings can accelerate particles to ultra-relativistic energies.  The
most commonly accepted scheme is first order Fermi acceleration
\citep[e.g.,][]{krymskii:77,blandford:78,bell:78,blandford:87}, which
typically produces a powerlaw distribution
\begin{equation}
	f(\gamma) = f_0 \gamma^{-s}
	\label{eq:powerlaw}
\end{equation}
with canonical index $s\sim2$.

For relativistic shocks the situation is not as clear cut.  Several
attempts have been made to solve the problem of particle acceleration at
relativistic shocks, mostly in the limit of test particle acceleration.  In
general, it is found that powerlaw distributions with somewhat steeper
spectra than in non-relativistic shocks can be produced
\citep{ellison:02,achterberg:01,kirk:87}.

It is certain, however, that acceleration of particles in the shock {\em
must} take place: Particles crossing the shock are by nature already
relativistic in the downstream rest frame, with a typical Lorentz factor of
$\Gamma_{\rm rel}$, the relative Lorentz factor between upstream and
downstream frames.  The particle distribution leaving the shock is thus
strongly anisotropic, and essentially mono-energetic.  The randomization of
this energy is then a question of the efficiency of the typical plasma
processes often assumed to be present in populations of relativistic
particles.

The simplest assumption is that the particle momenta are simply isotropized
behind the shock.  The shock acceleration kernel is then a delta function
and a cold upstream plasma will be transformed into a narrow but
relativistic energy distribution, the width $\sigma_{\Gamma}$ of which
should roughly be given by the Lorentz transformed width $\sigma$ of the
upstream energy distribution, $\sigma_{\Gamma} \sim \Gamma_{\rm rel}
\beta\, \sigma$, where $\Gamma_{\rm rel}$ is the relative Lorentz factor
between the upstream and downstream frames.  The mean particle energy will
be $\langle \gamma\,m\,c^2 \rangle \sim \Gamma_{\rm rel}\,m\,c^2 \sim
\Gamma_{\rm jet}\,m\,c^2$.

If scattering by downstream turbulence or particle interactions is
stronger, the particles can be thermalized, in which case a
Maxwell-Boltzmann distribution according to the relativistic
Rankine-Hugoniot jump conditions will be established.  The main
observational difference between these two cases will be the width of the
energy distribution (see Fig.~\ref{fig:sketch}).

If a significant fraction of the particles can perform multiple shock
crossings (which again hinges on effective scattering to return downstream
particles to the shock), we expect a powerlaw-type distribution to be
established.  It is reasonable to assume that the first time escape
fraction from the downstream region (i.e., the probability that a particle
which crossed the shock only once) is of order $\varP_{\rm esc} \lesssim
90\%$ \citep[e.g.,][]{achterberg:01}, which implies that most of the
particles will only cross the shock once \citep[note that the escape
probability in non-relativistic shocks is generally very small for fast
particles, though $\varP_{\rm esc}$ is of order unity for thermal
particles,][]{bell:78}.  The escape fraction in the upstream region is much
smaller and generally neglected in calculations.

The particles re-crossing the shock will pick up another factor of order
$\Gamma_{\rm rel}^2$ in energy gain \citep{vietri:95}, which implies that a
significant fraction of particles will be boosted to higher energies
($\gamma \geq \Gamma_{\rm rel}^3$).  This fraction of the particles will
contribute a significant amount of pressure to the post shock gas, which
will modify the shock structure accordingly.  Thus, the amount of energy
accessible to the bulk of the particles which cross the shock only once is
of the order $[1 + \Gamma_{\rm rel}^2 (1/\varP_{\rm esc} - 1)]^{-1} \sim
50\% - 90\%$.

The remaining fraction $1 - \varP_{\rm esc}$ of the particles will perform
true Fermi acceleration.  The low energy turnover of this distribution
should be located roughly at $\gamma \sim \Gamma_{\rm rel}^3$
(\citealt{vietri:95}) and subsequent shock crossings will produce features
at energies of $\Gamma_{\rm rel}^{2i + 1}$ (where $i$ enumerates the number
of shock crossings).  The similarity of this process to Compton
upscattering was already mentioned in Sect.~\ref{sec:spectrum_intro}.

The superposition of features from multiple shock crossing cycles will lead
to the formation of a powerlaw at high energies, very similar to the
powerlaw produced by optically thin inverse Compton scattering (where the
Lorentz transformations of the photon distribution to and from the particle
rest frame are replaced by Lorentz transformations to and from the upstream
fluid rest frame, assuming that scattering is strong enough to isotropize
the particle distribution.  The optical depth $\tau_{\rm IC}$ is replaced
by the return probability $1-\varP_{\rm esc}$.)  The shock powerlaw slope
$s_{\rm shock}$ is determined by $\Gamma_{\rm rel}$ and $\varP_{\rm esc}$
\citep[e.g.,][]{bell:78}:
\begin{equation}
	s_{\rm shock} \sim 1 - \frac{\ln{\left(1-\varP_{\rm
	esc}\right)}}{\ln{\left(\Gamma^2\right)}}
\end{equation}

Thus, observationally, diffusive acceleration will lead to the presence of
a second feature in the CR spectrum around $\sim \Gamma_{\rm jet}^3$.  This
situation is shown in Fig.~\ref{fig:sketch}, where the powerlaw spectral
component was plotted for $\Gamma_{\rm jet} = 5$.  The energy of this broad
peak coincides nicely with the peak in the excess CR proton component
required in the HEMN model \citep{strong:01b} to explain the $\gamma$-ray
{\em EGRET} excess above 1 GeV.

We note that \citet{achterberg:01} argue that in the absence of an
efficient scattering mechanism, the average energy gain per particle will
be restricted to a factor of order unity (rather than $\Gamma_{\rm rel}^2$)
for higher order shock crossings ($i \geq 2$), and that the low energy
turnover of the powerlaw component might thus be located at $\sim
\Gamma_{\rm rel}^2$ (rather than $\Gamma_{\rm rel}^3$).  The search for
additional features in the CR spectrum at energies $\Gamma^2$ and
$\Gamma^3$ might offer a potential way to test these predictions.

It is also possible that a population of relativistic protons already
exists upstream of the shock.  If the proton number density is equal to the
electron number density, the bulk of the particles and of the energy will
presumably be at the low energy end ($\gamma \sim 1$), otherwise the
estimates for kinetic energy flux in the jet would have to be increased
accordingly, increasing the impact on the Galactic CR spectrum as well (by
a factor of $\gamma_0$, the low energy cutoff of the distribution).

This population will be shifted to higher energies in the shock, and
possibly experience further Fermi type acceleration.  Thus, if a powerlaw
of relativistic protons exists prior to the terminal shock with lower
cutoff $\gamma_{0}$ and spectral index $s_{\rm jet}$, the terminal shock
should shift the lower cutoff roughly to $\Gamma_{\rm jet} \gamma_{0}$,
while the slope of the new powerlaw will be ${\rm Min}(s_{\rm jet}, s_{\rm
shock})$, where $s_{\rm shock}$ is the powerlaw slope produced in the
relativistic terminal shock.

While the particles produced in the shock must be relativistic at
injection, the dynamical evolution of the shocked gas can reduce their
energies significantly if adiabatic cooling is important before the
particles can escape the shock region (radiative cooling of the nucleon
distribution will be negligible).  The diffusion of particles out of radio
lobes and hot spots is a highly uncertain process and has not been studied
in the necessary detail to answer this important question. Rather than
discussing it at length, which would by far exceed the scope of this paper,
we decided to include a short discussion in Appendix
\ref{sec:appendix-diffusion}, where we show that adiabatic losses do indeed
pose a significant obstacle to particle escape (see also
Fig.~\ref{fig:sketch}).

\subsubsection{Electron acceleration}
CR protons do not emit any measurable amount of synchrotron radiation.
Inverse Compton scattering on protons is also very inefficient.  However,
if diffusive acceleration is operating efficiently in the working surface,
then the powerlaw electrons produced there should be detectable through
their radiative signature.  Tentative evidence of this powerlaw electron
component are the radio lobes in the microquasars 1E 1740-294 and GRS
1758-258.

The spectral index of the non-thermal emission from the lobes of 1E
1740-294 is $\alpha \sim 0.7 - 0.9$ \citep{mirabel:92}, corresponding to an
electron spectrum of slope of $s \sim 2.4 - 2.8$.  If the protons are
injected at the working surface with the same slope (this would require
that the electron spectrum is unaffected by radiative cooling and that the
powerlaw electrons are not just advected through the shock from upstream),
then the observed CR slope near the earth should be steepened by $\Delta s
= 1/2$ due to the energy dependence in the Galactic diffusion coefficient.
This would yield a proton powerlaw slope of $s\sim 2.9 - 3.3$, somewhat
steeper than the canonical CR slope of $s \sim 2.7$, though not
significantly.  

Part of the observed powerlaw electron distribution inside the tentative
radio lobes of 1E 1740-294 and GRS 1758-258 could also have been advected
from the jet.  In fact, is is unclear whether electrons can be accelerated
efficiently in a shock if protons are present, since the shock thickness
will be set by the proton Larmor radius, which will be much larger than the
electron Larmor radius \citep[e.g.,][]{achterberg:01}.  In this case, the
electrons will not experience a shock at all, more likely, they will be
accelerated adiabatically to a narrow component with energies of order
$\Gamma_{\rm jet}\,m_{\rm e}\,c^2$.  Such a component will not be
detectable through synchrotron radiation.

A detailed model of the energetic history of the particles
in the lobes of 1E 1749-294 would be required to answer these questions.

\section{Observational consequences}
\subsection{Predicted contribution to the Galactic CR spectrum}
Based on the picture laid out in Sect.~\ref{sec:acceleration}, we can try
to predict what might be observed in the CR spectrum due to the presence of
Galactic microquasars.

First, it is obvious that close enough to a powerful relativistic jet
source the locally observed CR spectrum will be completely
dominated by the CRs produced in the terminal shock of the jet.
However, it is clear that the powerlaw spectrum observed near earth is not
dominated by a narrow component of microquasar origin - the current
spectral limits rule out any contribution greater than a few percent.

In a simple isotropic diffusion picture, the CR energy density in the
environment of a continuously active source will fall roughly like the
inverse distance to the source $r^{-1}$ [see Eq.~(\ref{eq:crdensity})] for
large $r$ much larger than a particle mean free path, $r \gg \kappa/c \sim
0.2\,{\rm pc}\ H_{\rm kpc}^2/\tau_{15}$, and smaller than the Galactic disk
height, $r \lesssim H_{\rm disk} = 1\,{\rm kpc}\,H_{\rm kpc}$.

Given the observed CR energy density of $\sim 10^{-12}\ {\rm ergs\
cm^{-3}}$, we can estimate that the sphere of influence of a given source,
defined as the region inside which the source contributes more than 30\% of
the total measured CR power (at which level it would enter the
realm of detectability of by {\em AMS 02}) has a radius of order
\begin{equation} 
	R_{\rm 30\%} \sim 1\ {\rm kpc} \frac{L_{38.5}}{\kappa_{28.3}}.
\end{equation}
The time it takes the source to populate this sphere with CRs is
roughly
\begin{equation}
	\tau_{\rm 30\%} \sim 15\times 10^{6}\ {\rm yrs}
	\frac{L_{38.5}^2}{\kappa_{28.3}^3}
\end{equation}

Particles that reach the edge of the Galactic disk will be siphoned off
into the Galactic halo.  Thus the CR flux will diminish rapidly, roughly
exponentially, beyond a source distance of order the disc thickness, $r >
H$.  It would take source activity longer than $10^{8}\,{\rm yrs}$ to fill
the Galactic halo and the disk.

The well known microquasars mentioned above are all located much further
from the solar system than this limit.  However, if a source similar to,
say, GRS 1915+105 had been active in the solar neighborhood (inside about 1
kpc) within the last $\sim 10^{7}$ yrs, our local CR flux should show a
clear sign of the contribution from this source.

In this context it is important to mention that GRO J1655-40, V4641 Sgr,
Cyg X-3 (and also SS433) are known to be in high-mass X-ray binaries.
Their lifetimes are therefore expected to be short.  If such a relativistic
jet black hole binary was located in the Orion nebula region within the
past $10^{6}$ yrs, we should be able to detect a strong signal in the low
energy CR spectrum from this source alone.

Far enough away from any single source, an observer will measure the time
averaged contribution from all Galactic sources, washed out by CR diffusion
\citep[similar to the situation described in][]{strong:01a}.  Since sources
will likely follow a distribution of Lorentz factors of width $\Delta
\Gamma_{\rm jet}$, the observed signal will be smeared out over at least
that width.  Any intrinsic width of the produced CR spectrum will add to
this effect, as well as broadening effects like solar modulation and
scattering off of interstellar turbulence.

\begin{figure}[tpb]
\begin{center}
\resizebox{\columnwidth}{!}{\includegraphics{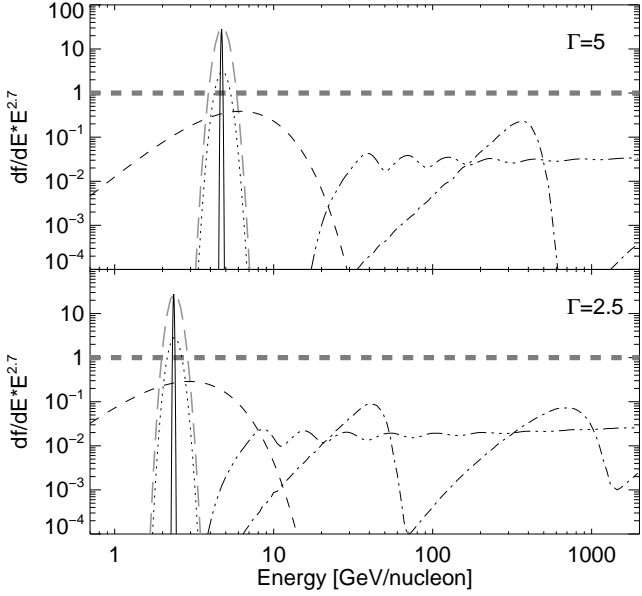}}
\end{center}
\caption{Toy model of the microquasar contribution to the CR spectrum, for
a {\em single} microquasar situated in a low mass X-ray binary, active for
$\tau \gtrsim 1.5\times\ 10^{7}\ {\rm yrs}$ on the level of $3\times
10^{38}\ {\rm ergs\ s^{-1}}$ (similar to GRS 1915+105), and at a distance
of $1\,{\rm kpc}$.  For simplicity, we assumed the source was operating
with uniform bulk Lorentz factor $\Gamma_{\rm jet} = 5$ (top) and
$\Gamma_{\rm jet} = 2.5$ (bottom).  The curves are normalized relative to
the measured differential Galactic CR background spectrum (thick grey
dashed line).  Shown are the same curves as in Fig.~\ref{fig:sketch}:
narrow feature for upstream temperature of $T \sim 7\times 10^{10}\,{\rm
K}$ ({\em dotted line}) and for $T \sim 7\times 10^{8}\,{\rm K}$ ({\em
solid line}), Maxwellian with $kT\sim \Gamma_{\rm jet}\,m_{\rm p}\,c^2/3$
(dashed line), multiply scattered component for efficient pitch angle
scattering ({\em dash-dotted line}) and for inefficient pitch angle
scattering ({\em dash-triple-dotted line}), and the relative contribution
of heavy elements for metallicity 10 times the solar value (long-dashed
grey curve), as seen in GRO\ J1655-40 and V4641\ Sgr.  Each spectral
component has been steepened by $E^{-1/2}$ to account for the energy
dependence of the diffusion coefficient.\label{fig:crspectrum}}
\end{figure}

In Fig.~\ref{fig:crspectrum} we have plotted possible contributions to the
CR proton spectrum from a single Galactic jet source.  Depending on how
much we have underestimated the power in Galactic jets and how much
adiabatic losses of particles trapped in adiabatically expanding shock will
suffer, we might over or underestimate the contribution.  Taking the figure
at face value, however, it seems likely that a contribution at the few
percent level can be expected in the energy region of a few GeV.

\subsection{Detectability of narrow features in the CR spectrum}
Detecting and positively identifying such a CR component will be a
formidable challenge.  The advent of the high sensitivity {\em AMS 02}
instrument and of the solar minimum might make it possible, however.  Given
the preliminary specifications of {\em AMS 02}, we can estimate the
detectability of spectral features such as produced by microquasars.  The
rigidity resolution (i.e., energy resolution) of the instrument is expected
to be around 2\% in the crucial range from 1 to 10 GeV, which will easily
be sufficient to identify and resolve even the narrowest feature in
Fig.~\ref{fig:crspectrum}.

For an effective area of order $0.4\,{\rm m^2\,sr}$, the expected total CR
proton count rate by {\em AMS 02} in the energy range from 1 to 10 GeV
should be of the order of $10^{3}\,{\rm s^{-1}}$.  At 2\% energy
resolution, this implies a detection rate of about $2\times 10^{8}\,{\rm
yr^{-1}\,bin^{-1}}$, with a relative Poisson-noise level of order
$10^{-4}$.  Calibration and other systematic errors will likely dominate
the statistics, however, these numbers are encouraging, and we expect that
a source at the few-percent level will be detectable with {\em AMS 02}.

The heavy element sensitivity of {\em AMS 02} will share similar
characteristics: for the same energy resolution and effective area, the
detection rates of Carbon and Iron, for example, should be of order
$4\times 10^{5}\,{\rm s^{-1}\,bin^{-1}}$ and $4\times 10^{4}\,{\rm
s^{-1}\,bin^{-1}}$ respectively.  Aside from {\em AMS 02}, signatures might
be detected by other instruments, and even existing data sets might contain
signals.  Identification would require scanning these data with high
spectral resolution.  Note that the effects of solar modulation will
broaden any narrow spectral component significantly.  Results by
\citealt{labrador:97} demonstrate that a line at $\sim 5\,{\rm GeV}$ will
be broadened by $\sim 1\,{\rm GeV}$, (less at higher energies) though this
effect will be reduced at solar minimum.

\subsection{Gamma ray emission from pion decay}
As the CRs produced in microquasars travel traverse the Galaxy, they will
encounter the cold ISM.  The interaction of a CR proton (by far the most
abundant and thus most energetic component of the CR spectrum) with a cold
ISM proton can lead to secondary particle production and to the emission of
gamma rays via several channels, the most important of which is $\pi^{0}$
decay.
\begin{figure}[tpb]
\begin{center}
\resizebox{\columnwidth}{!}{\includegraphics{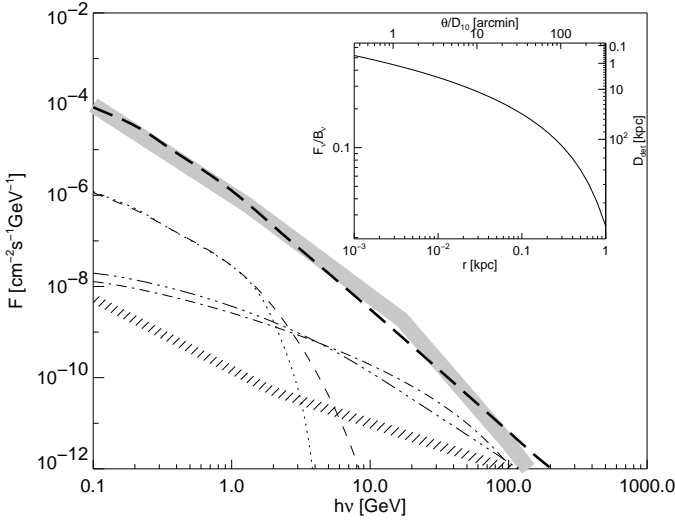}}
\end{center}
\caption{Toy model for the gamma ray signature produced in a microquasar CR
halo via pion decay (including $\pi^{0}$ decay and bremsstrahlung from
secondary electrons and positrons).  Curves were calculated assuming a
source active for over $\tau \gtrsim 15\times 10^{6}\,{\rm yrs}$ with an
average power of $L_{\rm kin} = 3\times 10^{38}\,{\rm ergs\,s^{-1}}$, at a
distance of 10 kpc, and for an ISM particle density of $1\,{\rm cm^{-3}}$,
assuming CRs are lost once they have reached the edge of the Galactic disk
at about 1 kpc distance from the source.  Labes according to
Fig.~\ref{fig:sketch}.  {\em dotted line}: Gamma ray signature from narrow
component of the microquasar halo; {\em dashed line}: from thermalized
component; {\em dash-dotted line}: from "powerlaw type component" in the
case of efficient scattering; {\em dash-triple-dotted line}: from "powerlaw
type component" in the case of inefficient scattering. {\em Thick grey
line}: {\em EGRET} diffuse gamma ray background at the position of GRS
1915+105 \citep{hunter:97}, assuming the same solid angle as subtended by
the source.  {\em Thick long dashed line}: background gamma ray emission
over the same solid angle modeled assuming the proton CR spectrum measured
near earth and an average ISM density of $1\,{\rm cm^{-3}}$.  Hatched line:
{\em GLAST} sensitivity.  Models were computed using the {\tt GALPROP}
routines by \citet{moskalenko:98}.  {\em Insert:} Angular dependence of the
source contribution to the gamma ray flux $F_{\nu}$, relative to the
background flux $B_{\nu}$ at a photon energy of $1\,{\rm GeV}$ [see
Eq.~(\ref{eq:geometry})].  The right Y-axis shows the source distance at
which the {\em GLAST} sensitivity is reached for the value of
$F_{\nu}/B_{\nu}$ shown in the curve, the top X-axis shows the angle
$\theta = r/D$ in units of arcmin$/D_{10}$ where $D_{10}$ is the source
distance in units of 10 kpc.
\label{fig:grspectrum}}
\end{figure}

Using the toy model presented in Fig.~\ref{fig:crspectrum}, we can estimate
how much gamma ray flux can be expected from the CR halo of a powerful
microquasar and compare it to the background flux from the Galaxy.  We
assume that the CRs diffuse away from the source until they reach the
Galactic halo, approximated as a zero pressure boundary condition at radius
$R \sim H_{\rm disk} = 1\,{\rm kpc}\,H_{\rm kps}$ (assuming spherical
symmetry for simplicity).  The result is shown in
Fig.~\ref{fig:grspectrum}.

Note that the gamma ray signal even for a source of average kinetic power
of $L_{\rm kin} = 3\times 10^{38}\,{\rm ergs\,s^{-1}}$ is small compared to
the background signal coming from the same solid angle ($\pi R^{2}$).
However, because the CR density increases towards the center of the source,
higher spatial resolution can improve the signal-to-noise ratio somewhat.
For a spherically symmetric cloud of CRs with luminosity $L$ and vanishing
pressure at the boundary $R \sim H_{\rm disk}$, the density follows
\begin{eqnarray}
	\lefteqn{n_{\rm CR}(r,\gamma) = \frac{f(\gamma)}{\int d\gamma'
	\gamma' m c^2 f(\gamma')}\frac{L}{4\pi \kappa}\left(\frac{1}{r} -
	\frac{1}{R}\right)} \\ & = & 6 \times 10^{-11}\,{\rm
	cm^{-3}}\frac{f(\gamma)}{\int d\gamma' \gamma' m c^2
	f(\gamma')}\frac{L_{38.5}\tau_{15}}{H_{\rm
	kpc}^2\,R_{kpc}}\left(\frac{R}{r} - 1\right) \nonumber 
	\label{eq:crdensity}
\end{eqnarray}

Obviously, then, the gamma ray surface brightness of such a source is
centrally peaked and thus the best observing strategy is to try to resolve
the source.  The photon flux $F_{\nu}$ from within an angular distance
$\theta \leq r/D$ from the source position (here, $D$ is the physical
source distance, while $\theta$ the on-sky angular distance from the source
position) is then proportional to
\begin{eqnarray}
	{F_{\nu}(\theta \leq \frac{r}{D})} & \sim & \frac{n_{\rm
	ISM}\,c}{4\pi\,D^2}\frac{L\,R^2}{3\,\kappa}\left(\frac{\int
	d\gamma' \sigma_{\nu}(\gamma') f(\gamma')}{\int d\gamma' \gamma' m
	c^2 f(\gamma')}\right) \nonumber \\ & & \times \left\{1 +
	3\,r^2\,\left[\ln{\left( 1 + \sqrt{1 - r^2}\right)} -
	\ln{\left(r\right)}\right]\right. \nonumber \\ & & \ \ \
	\left.-\left(1 + r^2\right)\sqrt{1 - r^2}\right\}
	\label{eq:geometry}
\end{eqnarray}
whereas the galactic background flux $F_{\nu,\,{\rm back}}$ for the same
solid angle in the small angle approximation is simply proportional to
$(r/D)^2$.  Thus, the source contribution relative to the background,
measured as $F_{\nu,\,{\rm source}}/F_{\nu\,{\rm back}}$, rises towards
lower $r$, which we have plotted in the insert in
Fig.~\ref{fig:grspectrum}.

\subsubsection{Sub-cosmic rays from sub-relativistic sources like SS433}
Objects like SS433 might be important sources of sub-cosmic rays in the
Galaxy: The cold thermal ions carried in the jet at $0.26\,c$ will be
accelerated to energies of order $E_{\rm ion} \equiv E_{30}\times 30\,{\rm
MeV}/{\rm nucleon}$.  These particles will suffer severe ionization and
Coulomb losses which will prevent them from traveling further than about
250 pc from the source.

However, they could act as a significant ionization source for the
surrounding medium: The ionization loss timescale for a particle with
energy $E = 30\,{\rm MeV}\,E_{30}$ is of order \citep{ginzburg:79}
\begin{equation}
	\tau_{\rm ion} \sim 2\times 10^{6}\,{\rm
	yrs}\,\frac{E_{30}^{3/2}}{n_{\rm ISM}},
\end{equation}
much shorter than the expected lifetime of the particles inside the
Galactic disk, i.e., the particles will deposit their energy in the
vicinity of the source, especially if the source is located within a
molecular cloud.

Furthermore, the excitation of nuclear $\gamma$-ray emission lines by
interaction of these sub-cosmic rays with interstellar heavy ions of C, O,
Fe, and other elements might be detectable by {\em INTEGRAL}.

\subsection{Cold electron-positron jets}

If the jet consists chiefly of relatively cold electron-positron plasma, and
if dissipation occurs mostly in the reverse shock, then the jet terminus
will produce relativistic electrons and positrons with energies of the
order of $\Gamma_{\rm jet}\,m_{\rm e}\,c^2 \sim 2.5\,{\rm MeV}$, which will
then begin to diffuse into the ISM.  Such positrons and electrons could
produce additional bremsstrahlung radiation at energies of a few hundreds of
keVs up to 2.5 MeV.  Much like mildly relativistic protons, these electrons
will contribute to the heating of the ISM due to the ionization losses, but
much more importantly for future observations might be the
electron-positron annihilation line at 511 keV.

For an integrated mechanical luminosity of $L_{\rm kin} = 3 \times
10^{38}\, {\rm ergs\,s^{-1}} \,L_{38.5}$ of the entire ensemble of Galactic
relativistic jets, the flux of positrons carried by jets is
\begin{equation}
	\dot{N}_{\rm e^{+}} = \frac{L_{\rm kin}}{2\,\langle \gamma
	\rangle\,\Gamma_{\rm jet}\,m_{\rm e}\,c^2} \sim 4\times
	10^{43}\,{\rm s^{-1}}\,\frac{L_{38.5}}{\Gamma_{5}\,\langle \gamma
	\rangle},
\end{equation}
where $\langle \gamma \rangle$ is the mean random Lorentz factor of the
electrons/positrons in the jet plasma.  Note that this estimate will be
independent of whether dissipation occurs mostly in the forward or reverse
shock.  Assuming that these positrons can diffuse out of the radio plasma
they are originally confined in, the estimated luminosity of the
annihilation line will be of the order of $L_{\rm ann} \sim L_{\rm
jet}/\langle \gamma \rangle$.

This is actually comparable to the total amount of positrons annihilating
in the Galaxy according to the observations of the e$^{+}$/e$^{-}$
annihilation line from {\em OSSE/GRO}, $\dot{N}_{\rm e^{+}, Gal} \sim
3\times 10^{43}\,{\rm s^{-1}}$ \citep{purcell:97}.  If Galactic jets are in
fact composed of electron-positron plasma, this measurement immediately
implies one of the following conclusions: {\em a)} either the mechanical
luminosity of these jets is not far above our relatively conservative
estimate of $3\times 10^{38}\,{\rm ergs\,s^{-1}}$, or {\em b)} the pair
plasma is not cold, i.e., $\langle \gamma \rangle \gg 1$, or {\em c)}
diffusion of particles across the magnetic boundary of the remnant jet
plasma is very inefficient, in which case many Galactic ``radio relics''
should exist, not unlike in the case of radio relics from radio loud AGNs
in the intracluster medium \citep[e.g.,][]{ensslin:98}.

The {\em Integral SPI} spectrometer and the {\em IBIS} imager would be able
to measure the increase in the annihilation line flux towards microquasars
located away from the Galactic center (where the background is highest)
like GRS1915+105, and to measure the line width if it could be detected.
These measurements could be very helpful in constraining the particle
content of relativistic Galactic jets.

\section{Conclusion}
We have argued that Galactic microquasars produce cosmic rays with energies
of a few GeV/nucleon.  Under preferable conditions (i.e., if particles in
the jet are cold and do not suffer significant adiabatic losses before
escaping the acceleration region) this component should be detectable as a
narrow peak with energy $\sim \Gamma_{\rm jet}\,m_{\rm p}\,c^2/{\rm
nucleuon}$ in the Galactic CR spectrum.  The superposition of several such
peaks would then reflect the distribution of Lorentz factors of Galactic
jet sources.

In addition, diffusive acceleration of particles might produce a powerlaw
distribution of particles with a low energy turnover at energies around
$\Gamma_{\rm jet}^2 m_{\rm p} c^2$ to $\Gamma_{\rm jet}^3 m_{\rm p} c^2$,
visible as an edge-like feature in the CR spectrum.

The locally measured contribution to the CR spectrum will be strongly
dominated by sources operating close by (within a distance of about one
Galactic disk height and within the past $10^7\,{\rm yrs}$), since at
distances much larger than that the contribution from a given microquasar
falls off exponentially.  We estimate the global energy content in the CR
component accelerated in Galactic relativistic jets to be at the 0.1\% to
10\% level of the total Galactic CR luminosity.

This CR contribution from Galactic relativistic jet sources might be
strongly overabundant in heavy elements, reflecting the composition of the
accretion disk where the jets are launched.  Thus, it is possible that the
chemical abundance measured in the GeV region (where we expect the
contribution from jets to show the strongest effect) will differ slightly
from the chemical composition at higher energies.

While signatures of the CR component from microquasars might already be
buried in existing data, the upcoming solar minimum and the launch of {\em
AMS 02} will offer ideal conditions to search for this component and to put
constraints on the microquasar activity in the nearby universe.

We suggested that the absence of any observable traces of microquasar
component in the cosmic ray proton spectrum could be used to argue in favor
of electron-positron jets.  For this case, we showed that existing {\em
OSSE/GRO} observations of the Galactic electron-positron annihilation rate
can be used to limit the power in cold electron-positron jets inside the
Galaxy to $L_{\pm} < 3\times 10^{38}\,{\rm ergs\,s^{-1}}$.

\begin{acknowledgements}We would like to thank Roger Blandford, Andrei
Bykov, Richard Mewaldt, Igor Moskalenko, and Vladimir Ptuskin for
insightful discussions and comments on the manuscript.  We would also like
to thank Andrew Strong for providing access to the {\tt GALPROP} code.  We
would like to thank the referee Rob Fender for helpful comments regarding
all aspects of the paper.  Rashid Sunyaev, as a Gordon Moore Scholar,
thanks Caltech for its hospitality during the work on this paper.  This
research has made use of the public GBI monitoring database hosted by NRAO.
\end{acknowledgements}

\begin{appendix}
\section{The particle spectrum produced in decelerating ejections}
\label{sec:appendix-slow-down}
It is straight forward to calculate the particle distribution produced by
single pass shock acceleration in a decelerating ejection in the
ultra-relativistic limit, assuming that one has knowledge of the single pass
shock acceleration kernel at a given shock velocity $\Gamma_{\rm rel}$.

Take an ejection of initial mass $M_0$ and Lorentz factor $\Gamma_0$, which
is sweeping up and shocking external matter.  The total energy of the
ejection and the swept up matter is
\begin{equation}
	E \sim \Gamma c^2 \left(M_0 + \int_0^N dN' \Gamma(N') m_{\rm p}
	\right) \label{eq:energy}
\end{equation}
where the integration is over the number $N$ of swept up particles with
mass $m_{\rm p}$ at a given ejection Lorentz factor $\Gamma$, and the extra
factor of $\Gamma$ inside the integral takes account of the shock
acceleration of the swept up particles.

Since $E$ is conserved, we can take the derivative of Eq.~(\ref{eq:energy})
with respect to $\Gamma$, and arrive at
\begin{equation}
	\frac{dN}{d\Gamma} \sim \frac{\left(\int_0^N dN' \Gamma(N') m_{\rm
	p} + M_0 \right)}{\Gamma^2 c^2} \sim \frac{E}{m_{\rm p} c^2}
	\Gamma^{-3}
\end{equation}
for $1 \ll \Gamma \leq \Gamma_0$.  For $\Gamma \sim 1$, the
non-relativistic corrections reduce the amount of energy available, leading
to a low energy cutoff at $\Gamma \sim 1$ and a high energy cutoff at
$\Gamma_{0}$.

To arrive at the observed particle distribution $dN/d\gamma$, this must be
convolved with the single shock acceleration kernel $\gamma(\Gamma)$,
however, for a narrow kernel, such as assumed in this paper, the powerlaw
approximation seems sufficient: $dN/d\gamma \propto \gamma^{-3}$.

\section{Cosmic ray diffusion and adiabatic cooling}
\label{sec:appendix-diffusion}
An important question is whether the particles produced in the shock
discussed in Sect.~\ref{sec:shock} can indeed diffuse out of the shock region,
in which case they will freely escape and propagate through the Galaxy
essentially with the energy obtained in the shock, or whether they are
trapped inside the shocked gas until it expands adiabatically after the
shock has passed and activity has ceased.  In the latter case, the
particles will lose a significant amount of energy to adiabatic expansion.

Following the discussion in Sect.~\ref{sec:shock}, we distinguish two
cases: dissipation in the forward and in the reverse shock.

The escape time of the particles out of the shock can be estimated as
\begin{equation}
	\tau \sim \frac{{R_{\rm shock}}^2}{\kappa}
\end{equation}
where $R_{\rm shock}$ is the typical size scale of the shock and $\kappa$
the relevant diffusion coefficient.

While the shocked ISM must still be magnetically connected with the
unshocked ISM, the jet plasma will be situated on field lines advected out
from the central engine, which are likely not connected with the ISM.  In
the forward shock, the relevant diffusion coefficient should then be taken
as $\kappa_{\rm forward} \sim \kappa_{\parallel}$, the diffusion
coefficient parallel to the mean magnetic field, while for the reverse
shock one has to consider diffusion across the magnetic boundary of the
contact discontinuity between shocked jet plasma and shocked ISM in
addition to diffusion to the contact discontinuity and away from it.

A lower limit on the diffusion time out of the shock is thus given by the
value for the forward shock, since the particles which have diffused out of
the reverse shock must, in addition, also propagate through the forward
shock.

\subsection{Forward shock:}
Using the simple approximate expression for $\kappa_{\parallel}$
\citep[e.g.,][]{kennel:66}, and assuming a Kolmogorov turbulence spectrum
for the magnetic field originating on scales of order the shock size
$R_{\rm shock}=l_{\rm jet}\,\theta=10^{15}\,{\rm cm}\,l_{16}\,\theta_{0.1}$
and containing a fraction $\epsilon \equiv {B_{\rm turb}^2}/{B_{\rm
tot}^2}$ of the total magnetic energy, the parallel diffusion coefficient
for a particle with energy $\gamma\,m\,c^2 \sim \Gamma_{\rm jet}\,m\,c^2$
can be written as
\begin{eqnarray}
	\kappa_{\parallel} & \sim & \frac{2}{\pi}\, v\, r_{\rm G}
	\frac{\nu_{\rm gyro}}{\nu_{\rm scatter}} \sim \frac{2\, r_{\rm G}\,
	c}{\pi}\frac{B^2}{3\,kB_{k}^2} \nonumber \\ & \sim & \frac{2 \,
	r_{\rm G}\, c}{3\pi\,\epsilon}\,\left(\frac{R_{\rm shock}}{r_{\rm
	G}}\right)^{2/3} \sim
	\frac{2\,c}{3\pi\,\epsilon}\left(\frac{\gamma\, m\,
	c^2}{e\,B}\,l_{\rm jet}^2\,\theta^2\right)^{1/3}.
\end{eqnarray}
Here, $\nu_{\rm scatter}$ is the scattering frequency for particles with
gyro radius $r_{\rm G}$ off magnetic turbulence with energy density
$k B_{\rm k}^2/8\pi$ and wave numbers of order $k \sim r_{\rm G}^{-1}$.
Scattering can also be induced by collective resonant interactions of
particles with the field, in which case the parameter $\epsilon$ denotes
the efficiency of this process.

Writing the shock area as $A_{\rm shock} = \pi R_{\rm shock}^2 = \pi \times
10^{30}\, {\rm cm^{2}}\ \theta_{0.1}^2\, l_{16}^2$ gives an approximate hot
spot pressure $p_{\rm shock}$ of
\begin{equation}
	p_{\rm shock} \sim \frac{L_{\rm kin}}{A_{\rm shock}\ c} = 3 \times
	10^{-3}\,{\rm ergs\ cm^{-3}}\, L_{38.5}\, l_{16}^{-2}\, \theta_{\rm
	0.1}^{-2}.  \label{eq:hs_pressure}
\end{equation}
If the magnetic field is at a fraction $\varphi$ of the equipartition
field, $\varphi \equiv B/B_{\rm eq}$, we have $B \equiv \varphi\,B_{\rm eq}
\sim 0.5\,{\rm G}\, \varphi\, L_{38.5}^{1/2}/l_{16}\, \theta_{0.1}$.

The {\em comoving} (i.e., measured in the frame of the shocked plasma)
limit to the proton escape time $\tau_{\parallel}'$ is then
\begin{eqnarray}
	\tau_{\parallel}' \gtrsim \frac{R_{\rm
	shock}^2}{\kappa_{\parallel}} & \sim & l_{\rm
	jet}\,\theta\,\frac{3\pi\,\epsilon\,\varphi^{1/3}}{2\,c}\,
	\left(\frac{e}{\gamma\,m\,c^2}\right)^{1/3}\, \left(\frac{L_{\rm
	jet}}{\pi\,c}\right)^{1/6} \nonumber \\ & \sim &
	4\times10^{7}\,{\rm
	s}\,\frac{l_{16}\,\theta_{0.1}\,\epsilon\,\varphi^{1/3}\,
	L_{38.5}^{1/6}} {\Gamma_{\rm jet}^{1/3}}.  \label{eq:diff_parallel}
\end{eqnarray}
If we write the Lorentz factor of the shocked ISM gas as $\Gamma'$, then
time dilation gives $\tau_{\parallel} = \Gamma'\,\tau_{\parallel}'$ in the
observer's frame.

If the turbulent velocity inside the region of interest is comparable to
the expansion velocity, and if large scale turbulence is present (which was
the underlying assumption in out estimate of $\epsilon$ above), then
turbulent transport could aid particle escape: In a simple mixing length
approach, the diffusion coefficient can be approximated by $\kappa_{\rm
turb} \sim 1/3\,v_{\rm turb}\,l_{\rm turb}$, where $v_{\rm turb}$ is the
characteristic turbulent velocity and $l_{\rm turb}$ the scale length of
the largest scale turbulence.  The turbulent transport time is then
\begin{equation}
	\tau_{\rm turb} \sim \frac{3 R_{\rm shock}^2}{v_{\rm turb}\,l_{\rm
	t}} \sim 10^{7}\,{\rm s}\,\frac{\dot{R}_{\rm shock}}{v_{\rm
	t}}\frac{0.1\,R_{\rm shock}}{l_{\rm turb}}
\end{equation}

If the shock region is expanding roughly ballistically as it propagates,
the adiabatic loss time scale measured in the {\em observer's frame} is
\begin{equation}
	\tau_{\rm ad} = \frac{\gamma}{\dot{\gamma}} = 4\frac{p}{\dot{p}} =
	\frac{R_{\rm shock}}{\dot{R}_{\rm shock}} \sim 3\times 10^{5}\,{\rm
	s}\ l_{16}\,\left(\frac{c}{\dot{l}}\right).
\end{equation}

If the forward shock is relativistic, we need to take time dilation into
account when comparing $\tau_{\rm ad}$ and $\tau_{\parallel}$.  It follows
that the adiabatic loss time scale $\tau_{\rm ad}$ is longer than the
escape time only if
\begin{equation}
	\epsilon\,\varphi\, L_{38.5}^{1/6}\,
	\theta_{0.1}\,\frac{\dot{l}_{\rm jet}}{c}\,\Gamma_{5}^{2/3} < 4
	\times 10^{-3}.
\end{equation}
Since both $\epsilon$ and $\varphi$ might be significantly smaller than
$10^{-5}$, it is not implausible that $\tau_{\parallel} \gtrsim \tau_{\rm
ad}$.  In the case of strong turbulent transport, this condition simplifies
to
\begin{equation}
	v_{\rm turb}\,l_{\rm turb} > 3 R_{\rm shock} \dot{R}_{\rm shock}.
\end{equation}

If the external pressure is parameterized as $p_{\rm ISM} \equiv
10^{-11}\,{\rm ergs\ cm^{-3}}\, p_{-11}$, the maximum reduction in particle
energy possible through adiabatic losses is roughly
\begin{eqnarray}
	\left(\frac{p_{\rm ISM}}{p_{\rm shock}}\right)^{1/4} & \sim &
	10^{-2} \left(\frac{p_{-11}\, l_{16}^2\,
	\theta_{0.1}^2}{L_{38.5}}\right)^{1/4} \\ & \lesssim &
	\frac{\langle\gamma - 1\rangle_{\rm min}}{\langle\gamma -
	1\rangle_{\rm shock}} < \left(\frac{p_{\rm ISM}}{p_{\rm
	s}}\right)^{2/5} \nonumber \\ & & \hspace{48pt} \sim 6\times
	10^{-4} \left(\frac{p_{-11}\, l_{16}^2\,
	\theta_{0.1}^2}{L_{38.5}}\right)^{2/5}\nonumber,
\end{eqnarray}
where the left hand side of the inequality corresponds to a relativistic
equation of state in the hot spot and the right hand side to a
non-relativistic equation of state.  This, of course, means that a large
fraction of the energy injected originally into CRs could be lost
adiabatically, reducing the amount of energy available (estimated in
Sect.~2.1) by up to a factor of $\sim 1000$.

\subsection{Reverse shock:}
In jets where dissipation occurs mainly in the reverse shock, the jet
geometry is similar to AGN jets (i.e., the jets are effectively reflected
by the ISM, thus inflating a cocoon with spent jet fuel).

Particles accelerated in the shock will eventually be advected out of the
shock region and into the cocoon (see Fig.~\ref{fig:cartoon}).  The
timescale for this process is
\begin{equation}
	\tau_{\rm advect} \sim \frac{R_{\rm shock}}{v_{\rm advect}} \sim 3
	\times 10^{4}\,{\rm s}\,l_{16}\,\theta_{0.1}\,\left(\frac{c}{v_{\rm
	advect}}\right),
\end{equation}
where $v_{\rm advect} \lesssim c/3$, since the limiting downstream velocity
in a strong relativistic shock is $c/3$.

The diffusion time towards the contact discontinuity is given by
Eq.\,(\ref{eq:diff_parallel}).  In order to enter the forward shock, they
have to propagate across the magnetic boundary at the contact
discontinuity, which introduces an additional cross-field diffusion term.
The contact discontinuity will have a typical thickness of the order of the
Larmor radius $r_{\rm G}$ of the particles, thus the particles will have to
traverse a region of size $r_{\rm G}$ perpendicular to the field in order
to cross the contact discontinuity (this approximation is valid as long as
the parallel diffusion time over one coherence length of the field is
longer than the perpendicular diffusion time across one Larmor radius,
otherwise the perpendicular diffusion time across one coherence length
should be used).  The lower limit to the diffusion time across the field is
then $\tau_{\perp} \gtrsim \frac{r_{\rm G}^2}{\kappa_{\perp}}$.

The perpendicular diffusion coefficient $\kappa_{\perp}$ can be
approximated as $\kappa_{\parallel}\,(k B_{\rm k}^2/B^2)^2$
\citep[e.g.,][]{parker:65}, i.e.,
\begin{eqnarray}
	\tau_{\perp} & \sim & \tau_{\parallel}\, \left(\frac{r_{\rm
	G}}{R_{\rm shock}}\right)^2\,\left(\frac{\kappa_{\parallel}}
	{\kappa_{\perp}}\right) \sim \tau_{\parallel}\,\left(\frac{r_{\rm
	G}}{R_{\rm shock}}\right)^{2}\,\left(\frac{B^2}{kB_{k}^2}\right)^2
	\nonumber \\ & \sim & \frac{3\pi}{2\,c\,\epsilon}\,R_{\rm
	s}^{2/3}\,r_{\rm G}^{1/3} \sim 160\,{\rm
	s}\,\frac{\gamma^{1/3}\,l_{16}\,\theta_{0.1}}
	{\epsilon\,L_{38.5}^{1/6}\,\varphi^{1/3}} \label{eq:diff_perp}
\end{eqnarray}

The total diffusion time out of the reverse shock should then be of the
order of $\tau \sim \tau_{\parallel} + \tau_{\perp}$, which has a minimum
when $\left(\epsilon\,\varphi^{1/3}\,L_{38.5}^{1/6}/\Gamma_{\rm
jet}^{1/3}\right)_{\rm min} \sim 3\times 10^{-3}$, with $\tau_{\rm min}
\sim 2\times 10^{5}\,l_{16}\,\theta_{0.1}$.  Thus, even in the optimal
case, the diffusion time out of the reverse shock is longer than the
advection time unless $v_{\rm advect} \lesssim 0.1\,c$.

Thus, it is rather likely that the bulk of the particles are advected out
of the shock and into the cocoon.  The pressure driven expansion of a
cocoon can be approximated by a simple spherically symmetric model:
Dimensional analysis suggests that the size of the cocoon $R_{\rm c}$
follows a simple scaling \citep{castor:75}
\begin{equation}
	R_{\rm c} \sim \left(\frac{L\,t^3}{\rho_{\rm ISM}}\right)^{1/5}.
\end{equation}

Using this scaling to obtain order of magnitude estimates of the conditions
inside the cocoon, we can write the cocoon pressure $p_{\rm c}$ as
\begin{eqnarray}
	p_{\rm c} & \sim & \frac{1}{2}\,\rho_{\rm ISM} \dot{R}_{\rm c}^2
	\nonumber \\ & \sim & 2\times 10^{-5}\,{\rm ergs\,
	cm^{-3}}\,R_{16}^{-4/3}\,L_{38.5}^{2/3}\,n_{\rm ISM}^{1/3},
	\label{eq:cocoon_pressure}
\end{eqnarray}
where $R_{16} = R_{\rm c}/10^{16}\,{\rm cm} \sim l_{\rm 16}$ is the
cocoon radius.  The magnetic field is
\begin{equation}
	B_{\rm c} \sim 3\times 10^{-2}\,{\rm
	G}\,\varphi\,R_{16}^{-2/3}\,L_{38.5}^{1/3}\,n_{\rm ISM}^{1/6}
\end{equation}

Comparison of Eq.\,(\ref{eq:cocoon_pressure}) with
Eq.\,(\ref{eq:hs_pressure}) shows that a (relativistic) particle advected
out of the shock into the cocoon will suffer adiabatic losses of the order
\begin{equation}
	\frac{\langle \gamma - 1 \rangle_{\rm c}}{\langle \gamma - 1
	\rangle_{\rm shock}} \sim \left(\frac{p_{\rm c}}{p_{\rm
	s}}\right)^{1/4} \sim 0.3
	\left(\frac{\theta_{0.1}^{2}\,R_{16}^{2/3}\,n_{\rm
	ISM}^{1/3}}{L_{38.5}^{1/3}}\right)^{1/4}
\end{equation}
which is only weakly dependent on the source parameters.

Once the particles are inside the cocoon, the escape time is again given by
$\tau \sim \tau_{\parallel} + \tau_{\perp}$ with the expressions for
$\tau_{\parallel}$ and $\tau_{\perp}$ from Eq.\,(\ref{eq:diff_parallel})
and Eq.\,(\ref{eq:diff_perp}), though with different values:
\begin{equation}
	\tau_{\parallel} \sim 7\times 10^{8}\,{\rm
	s}\,R_{16}\,\left[\epsilon\,\varphi^{1/3}\,
	R_{16}^{1/9}\,\Gamma^{-1/3}\, L_{38.5}^{1/9}\,n_{\rm
	ISM}^{1/18}\right]
\end{equation}
and
\begin{equation}
	\tau_{\perp} \sim 4.5\times 10^{3}\,{\rm
	s}\,R_{16}\,\left[\epsilon\,\varphi^{1/3}\,
	R_{16}^{1/9}\,\Gamma^{-1/3}\, L_{38.5}^{1/9}\,n_{\rm
	ISM}^{1/18}\right]^{-1}
\end{equation}

The adiabatic loss timescale is simply
\begin{equation}
	\tau_{\rm ad} = 4\frac{p_{\rm c}}{\dot{p}_{\rm c}} \sim
	\frac{R_{\rm c}}{\dot{R_{\rm c}}} \sim 7\times 10^{5}\,{\rm
	s}\,R_{16}^{5/3}\,n_{\rm ISM}^{1/3}\,L_{38.5}^{-1/3}.
\end{equation}
Since this is proportional to $R^{5/3}$, while $T_{\parallel} \propto
R^{10/9}$ and $\tau_{\perp} \propto R^{8/9}$, the particles are more likely
to escape out of large, old cocoons than out of small, young ones.

\subsection{Summary}
The diffusion of CRs out of radio lobes is clearly an important problem not
only in the context of CR emission from microquasars, but also for
extragalactic radio sources and their contribution to the CR flux in
clusters of galaxies.  Since the current level of understanding is still
very rudimentary, further study of this aspect is necessary.  Given the
large uncertainties in these estimates, and given that we tried to provide
conservative estimates whenever possible, we feel that there is a good
chance that the mechanisms outlined in this paper will work under realistic
circumstances and that a measurable CR contribution from microquasars can
be expected.
\end{appendix}

\newpage
\end{document}